\documentclass{IEEEtaes}

\usepackage{color,array,amsthm}
\usepackage{graphicx}
\usepackage{url}
\usepackage[ruled]{algorithm2e}
\usepackage{amsmath,graphicx}
\graphicspath{{./figures/}{./figures/}}
\usepackage{import}
\usepackage[font=small]{caption}
\usepackage{subfig}
\usepackage{paralist}
\usepackage{empheq}
\usepackage{orcidlink}
\usepackage{float}
\usepackage{svg}
\usepackage{hyperref}
\usepackage{cite}
\usepackage{siunitx}
\DeclareSIUnit{\rad}{rad}
\usepackage{xfrac}
\usepackage{amsfonts}
\usepackage{amsmath}
\usepackage{amssymb}
\usepackage{mathrsfs}
\usepackage{dsfont}
\usepackage{url}
\usepackage{xargs}
\usepackage{color}
\definecolor{tcolor}{RGB}{0,60,104}
\usepackage{graphicx}

\jvol{XX}
\jnum{XX}
\jmonth{XXXXX}
\paper{1234567}
\pubyear{2020}
\doiinfo{TAES.2020.Doi Number}

\usepackage[shortcuts, acronym]{glossaries}
\glsdisablehyper

\newcommand{\tp}{\mathrm{T}}
\newcommand{\mat}[1]{{\ensuremath{{\mathbf{#1}}}}}

\DeclareMathOperator{\diag}{diag}

\newacronym{ekf}{EKF}{Extended Kalman Filter}
\newacronym{eot}{EOT}{Extended Object Tracking}
\newacronym{gwd}{GWD}{Gaussian Wasserstein Distance}
\newacronym{gws}{GWS}{Gaussian Wasserstein Score}
\newacronym{radar}{radar}{Radio Detection And Ranging}
\newacronym{lidar}{lidar}{Light Detection And Ranging}
\newacronym{ukf}{UKF}{Unscented Kalman Filter}
\newacronym{mse}{MSE}{Mean Squared Error}
\newacronym{memekf}{{MEM-EKF*}}{Multiplicative Error Model Extended Kalman Filter*}
\newacronym{iou}{IoU}{Intersection over Union}
\newacronym{rm}{RM}{Random Matrix}
\newacronym{rbpf}{RBPF}{Rao-Blackwellized Particle Filter}
\newacronym{evd}{EVD}{Eigenvalue Decomposition}
\newacronym{lomem}{L:OMEM}{Lambda:Omicron Multiplicative Error Model}
\newacronym{mem}{MEM}{Multiplicative Error Model}
\newacronym{memrbpf}{MEM-RBPF}{Multiplicative Error Model Rao-Blackwellized Particle Filter}
\newacronym{vbrm}{VBRM}{Variational Bayes Random Matrix}
\newacronym{rmse}{RMSE}{Root Mean Square Error}
\newacronym{memqkf}{MEM-QKF}{Multiplicative Error Model Quadratic Kalman Filter}
\newacronym{memeqkf}{MEM-eQKF}{Multiplicative Error Model Exponential-form Quadratic Kalman Filter}
\newacronym{imm}{IMM}{Interacting Multiple Model}
\newacronym{bev}{BEV}{Bird's Eye View}
\newacronym{memeif}{MEM-EIF}{Multiplicative Error Model Extended Information Filter}

\newcommand{\rot}[1]{\text{rot}\left(#1\right)} 
\newcommand\EX{\mathbb{{E}}}
\DeclareMathOperator{\var}{var}
\DeclareMathOperator{\vect}{vect}
\DeclareMathOperator{\cov}{cov}
\DeclareMathOperator{\trace}{tr}

\newcommand\scalingfactor{c}
\newcommand\generalcov{\mat{{P}}}
\newcommand\kinstate{\mat{{m}}}
\newcommand\kincov{\generalcov^{\kinstate}}
\newcommand\fullstate{\mat{{x}}}
\newcommand\fullcov{\generalcov^{\fullcov}}
\newcommand\meascov{\mat{{R}}}

\newcommand\semiaxes{\mat{{l}}}
\newcommand\axisstate{\boldsymbol{\gamma}}

\begin{document}

\title{Decoupled Quadratic Kalman Filter for Elliptical Extended Object Tracking with Log-normal Axis Modeling}

\author{SIMON STEUERNAGEL}
\affil{University of Göttingen, Göttingen, Germany} 

\author{MARCUS BAUM}
\affil{University of Göttingen, Göttingen, Germany}

\receiveddate{Manuscript received XXXXX 00, 0000; revised XXXXX 00, 0000; accepted XXXXX 00, 0000.}

\corresp{{\itshape (Corresponding author: S. Steuernagel)}.}

\authoraddress{
Simon Steuernagel and Marcus Baum are with the Institute of Computer Science, University of Göttingen, Göttingen, 37077 Germany. (e-mail: \{simon.steuernagel, marcus.baum\}@cs.uni-goettingen.de).
\\This work has been submitted to the IEEE for possible publication. Copyright may be transferred without notice, after which this version may no longer be accessible.
}

\supplementary{Color versions of one or more of the figures in this article are available online at \href{http://ieeexplore.ieee.org}{http://ieeexplore.ieee.org}.}

\markboth{STEUERNAGEL ET AL.}{DECOUPLED EXTENDED OBJECT TRACKING}
\maketitle

\begin{abstract}
Extended object tracking involves estimating both the physical extent and kinematic parameters of a target object, where typically multiple measurements are observed per time step.
In this article, we propose a deterministic closed-form elliptical extended object tracker, based on decoupling of the kinematics, orientation, and axis lengths. 
By disregarding potential correlations between these state components, fewer approximations are required for the individual estimators than for an overall joint solution.
This also enables a log-normal representation of the semi-axis lengths, i.e., opposed to related approaches, only positive axis lengths have support.
The resulting algorithm outperforms existing algorithms, reaching the accuracy of sampling-based procedures. 
Additionally, a batch-based variant is introduced, yielding highly efficient computation while outperforming all comparable state-of-the-art algorithms.
This is validated both by a simulation study using common models from literature, as well as an extensive quantitative evaluation on real automotive radar data.
\end{abstract}


\glsresetall
\glsunset{memrbpf}
\glsunset{memekf}
\glsunset{memeif}

\newpage
\section{INTRODUCTION}~\label{sec:introduction}
\begin{figure}[t]
    \centering
    \includegraphics[width=\columnwidth]{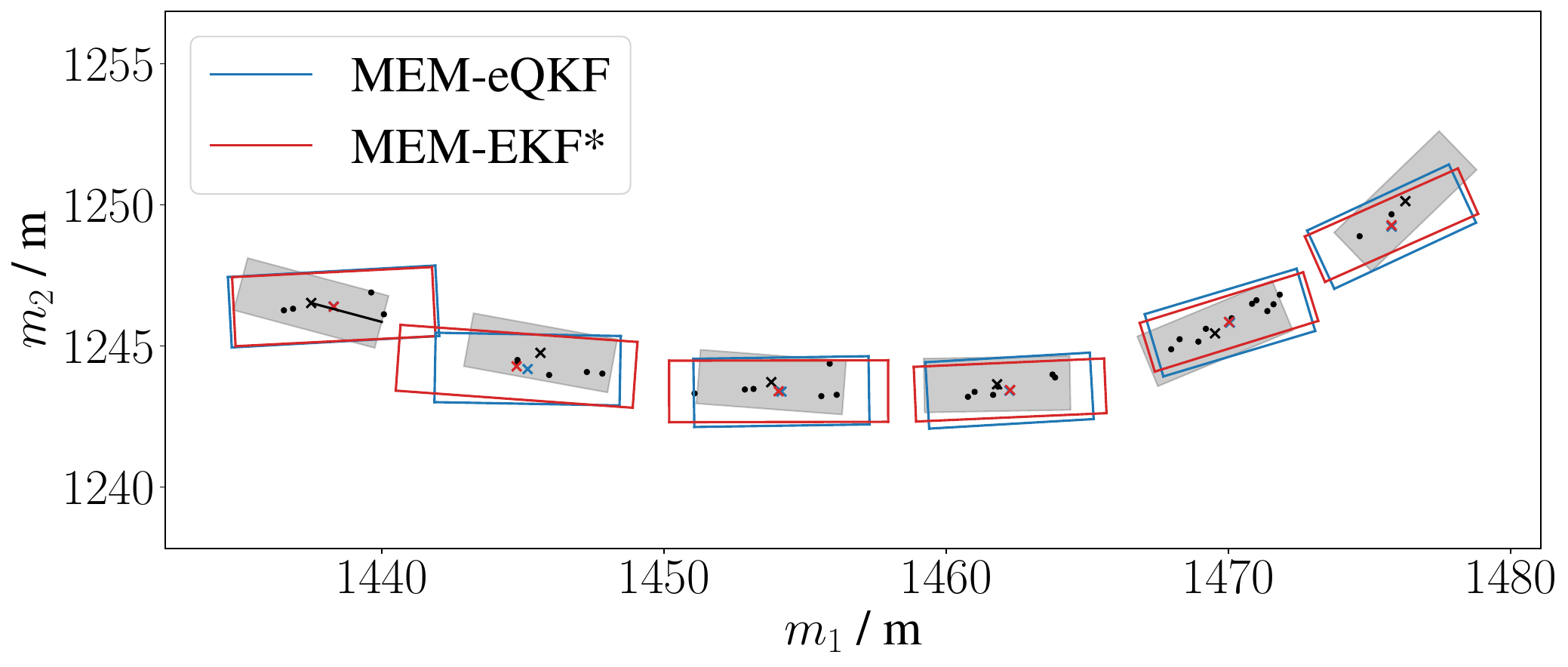}
    \caption{
    Example tracking results using automotive radar data from the nuScenes data set~\cite{caesar2020nuscenes}. The proposed \acrshort{memeqkf} (blue) converges noticeably faster to a good estimate compared to the \acrshort{memekf}~\cite{yang2019tracking} (red), starting from the same prior (leftmost estimate). Measurements are shown as black dots, and the ground truth annotation in gray.
    }
    \label{fig:radar_example}
\end{figure}

T{\scshape racking} is a key technology for a wide variety of perception systems.
It is employed using a range of sensor modalities, e.g., cameras, radars, or lidars. Each of these prompts its own set of challenges. 
Traditionally, tracked objects were often modeled as point targets, i.e., only as their position and further kinematic parameters such as velocity.
However, for many applications, the physical extent, i.e., shape, of the target is of interest, for example, in order to classify the object or to perform collision avoidance.
In sparse point cloud data as produced by, e.g., radar, the extent cannot be easily determined using a single scan of data. Instead, information needs to be integrated over time, commonly done in a Bayesian filtering framework. 
This is referred to as \ac{eot}~\cite{granstroem_baum_2022}, an example for which can be seen in Fig.~\ref{fig:radar_example}.
To track multiple extended objects~\cite{wieneke2012pmht, granstrom2016gamma, yang2020marginal}, the state of each object must be estimated from the data. This is what this article is concerned with, i.e., the estimation of the kinematics and the extent of each individual object.

A variety of models for the target shape have been proposed. Star-convex shapes~\cite{baum2014extended, wahlstrom2015extended} offer large flexibility in modeling the target extent. 
Even more flexibility is achieved with fully free-form approaches, such as the one recently proposed in~\cite{kumru2024tracking}.
However, one of the most common approaches in literature is to model the object extent as an ellipse~\cite{feldmann2010tracking, yang2019tracking, govaers2019independent, tesori2023lomem}. The reduced size of the resulting parameter space ensures a more robust fit in sparse and noisy data, as often encountered in \ac{eot}.
Furthermore, the parameterizations of an ellipse and a rectangle are identical, meaning the models can often be easily transferred between the two representations.

Different concepts for tackling the problem of elliptical \ac{eot} have been proposed.
By modeling the shape of the object as a \ac{rm}, the shape estimation can be derived in a Bayesian fashion based on the scaled sample covariance of the measurements~\cite{koch2008bayesian, feldmann2010tracking}. A conceptual downside of this approach is that the overall uncertainty of all shape parameters, i.e., both axes and the orientation, is represented by a single scalar value.
By using variational Bayes, explicit consideration of the orientation based on the \ac{rm} framework has been explored in~\cite{tuncer2021random}. 

An alternative is to model the extent based on a multiplicative factor in the measurement equation, which has been proposed in~\cite{baum2012modeling}. The resulting \ac{mem} forms the basis of several trackers~\cite{yang2019tracking, tesori2023lomem, steuernagel2025extended}. Recently,~\cite{zhao2025adaptive} proposed a reinforcement learning-based tracker using it.
In our previous work~\cite{steuernagel2025extended}, a \ac{rbpf} for elliptical \ac{eot} using the \ac{mem} has been derived. Numerical results show the accuracy of the tracker. However, by nature of being particle-based, it can be categorically unsuitable for specific applications, such as certain safety-critical ones.
A closed-form, deterministic variant can hence be favorable.

Furthermore, the discussed existing \ac{mem}-based approaches all model the state as Gaussian, meaning that negative semi-axis lengths are supported by the underlying probabilistic model, despite not having any physical meaning. In practice, if the variance of the Gaussian is large enough with respect to the mean, the size estimate can ``collapse''.
It would be favorable to make use of a probabilistic representation that accounts for the physical properties. In this article, we propose to handle this by representing the axis in log-normal form.

\subsection{Contribution}
\begin{figure}[t]
    \centering
    \includegraphics[width=\columnwidth]{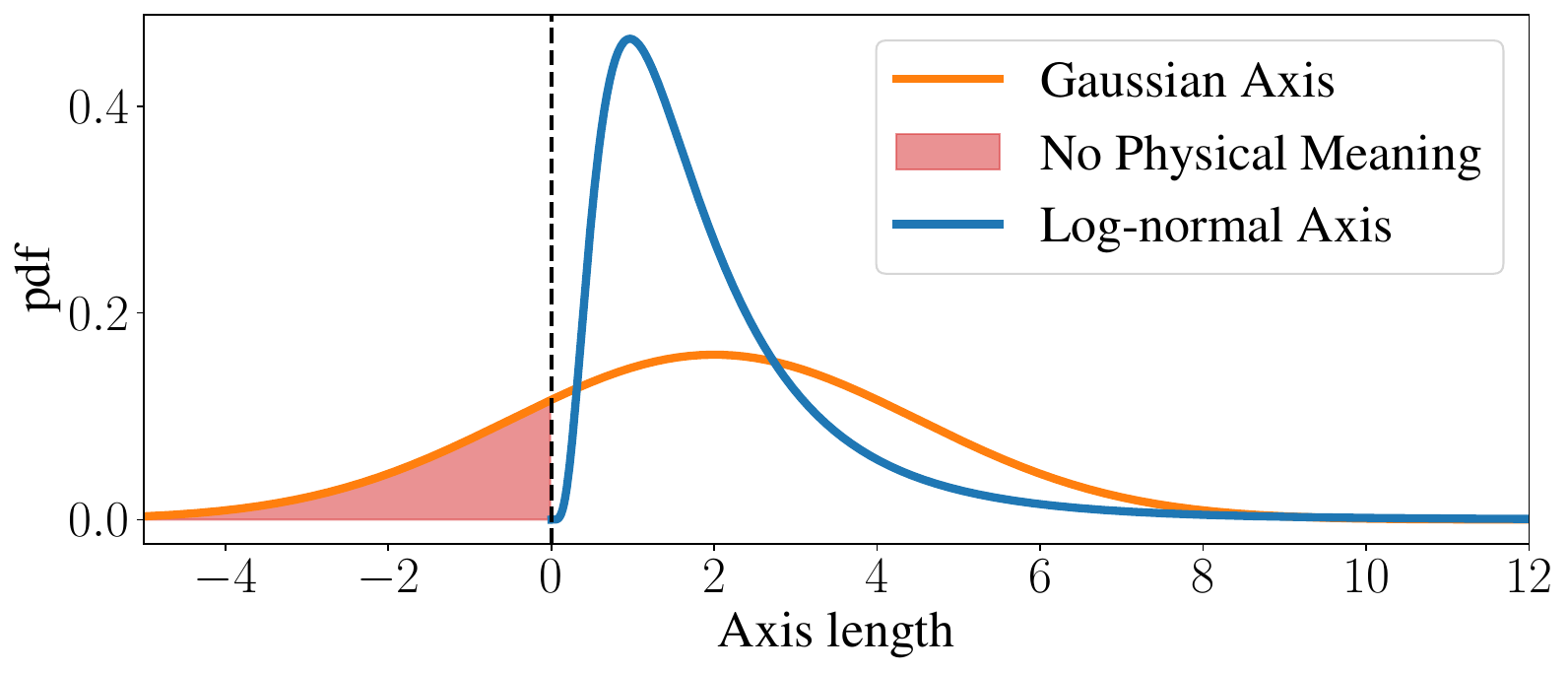}
    \caption{
    Comparison between Gaussian and log-normal representations of the same axis length. The red shaded area indicates support of the Gaussian distribution for physically impossible negative axis length.
    }
    \label{fig:lognormal}
\end{figure}

The key contribution of this article is a novel closed-form extended object tracker that yields significant improvements over the state-of-the-art.
We propose a decoupled approach to the problem, with individual estimators for kinematics, orientation, and semi-axis lengths. 
This approach allows foregoing specific approximation necessary for state-of-the-art approaches. Crucially, it allows derivation of the closed-form size estimator in log-normal form instead of requiring a Gaussian distribution for the semi-axes. Therefore, physically impossible negative semi-axes are not modeled, see Fig.~\ref{fig:lognormal}.
The algorithm is shown to be overall significantly more effective compared to related reference algorithms, while also being robust to several potential corner-cases.

In addition to a sequential method which explicitly iterates over measurements, we derive a faster batch-based variant also using the log-normal axis representation, which performs favorable compared to corresponding state-of-the-art filters.
Linearization allows the derivation this closed-form batch-based approach using the same moments that are shown to be effective approximations already by the sequential variant.

Results are validated on a simulation study covering a variety of measurement settings, consistently observing improved results.
Furthermore, a large-scale evaluation on real-world automotive data is conducted, comparing many model-based trackers from literature to the proposed system, further confirming the results.
While previous works have compared such trackers on individual real-world scenarios~\cite{cao2021automotive} or on a larger number of sequences but for learning-based algorithms~\cite{xia2021learning} or individual methods~\cite{kaulbersch2020assymetric}, to the best of our knowledge, the model-based trackers evaluated here have never been compared on real-world data at this scale.

In summary, the proposed algorithm sets a new standard for closed-form elliptical \ac{eot} filters. Closely related to the \ac{memekf}~\cite{yang2019tracking} and its batch-based variant~\cite{gramsch2024batch}, the two variants of the proposed algorithm supersede these algorithms in all conducted experiments on both simulated and real-world data. 
Importantly, we eliminate the Gaussian axis model used in these algorithms with one that is consistent with the underpinning physical constraints.

\subsection{Conference Paper}
This work is a further development and extension of the conference paper~\cite{steuernagel2025decoupled}, where the decoupled framework was first introduced.
The conference version still used the Gaussian semi-axis representation common in literature. Here, we instead derive the size estimation for the physically grounded log-normal representation.
While the empirical performance of the two variants is close, this solves an important theoretical weakness.

Furthermore, beyond the conference version, we introduce a computationally more efficient batch-based variant.
Additionally, this version includes considerably extended simulation studies including several potential corner-cases as well as evaluation on real-world radar sequences.
Finally, the main body of the article has been revised and significantly expanded.

\subsection{Notation}
Throughout this article, the superscript $i$ denotes the measurement index, corresponding to processing measurement $\mat{z}^i_k, i=1\dots M_k$. For iterative updates of the state, the index $i=0$ corresponds to the prior, and $i=M_k$ to the posterior. A superscript in brackets selects the indexed element from a vector or matrix.  $\rot{\cdot}$ denotes a standard two-dimensional rotation matrix, and $\diag(\cdot)$ constructs a diagonal matrix from the parameters. 

\subsection{Article Structure}
In the following, Section~\ref{sec:related_works} will extensively discuss existing works on the topic at hand, followed by Section~\ref{sec:problem_setting} where the problem setting of the article will be defined.
Next, the proposed approach will be illustrated in Section~\ref{sec:methodology} for the sequential and Section~\ref{sec:batch} for the batch-based variant.
Afterwards, analysis and evaluation of the results will be presented for the simulation study in Section~\ref{sec:simulation} and subsequently for the experiments conducted on real-world radar data in Section~\ref{sec:nuscenes}.
The article is then concluded in Section~\ref{sec:conclusion}.

\section{RELATED WORK}~\label{sec:related_works}
A variety of previous works have tackled the problem of (elliptical) extended object tracking. In the following, these will be summarized and contrasted with the approach presented in this article. 

Our proposed algorithm is based on the \ac{mem}, which was introduced in~\cite{baum2012modeling}. Subsequently, other approaches also based on this model will be discussed.
The most prominent \ac{mem}-based tracker is the \ac{memekf}, proposed in~\cite{yang2019tracking}. Measurements are processed sequentially by an \ac{ekf} which is specifically tailored to the problem at hand. The resulting filter can be computed in closed form.
With a few additional approximations, it is possible to process a batch of measurements simultaneously, as shown in~\cite{gramsch2024batch}. 
A multi-object tracker for extended objects based on~\cite{yang2019tracking} has been proposed in~\cite{yang2020marginal}.
In summary, the \ac{memekf} is closely related to the proposed method, and our method could be viewed as its successor.

A different filter, working under the assumption that the orientation can be approximated as the direction of the velocity vector was proposed in~\cite{tesori2023lomem}.
In this work, however, we are concerned with the general problem of estimating not only the kinematics and axes, but also the orientation.
This problem setting is also tackled by the \ac{rbpf} derived in~\cite{steuernagel2025extended}, which yields significantly improved results compared to the \ac{ekf} from~\cite{yang2019tracking}. This is achieved by accurate orientation estimation due to sampling, along with improved axis estimation by making use of the sampled orientation.
However, as it is not a closed-form method, a stark conceptual difference to the method proposed here can be seen.

In comparison to all these methods~\cite{yang2019tracking, tesori2023lomem, steuernagel2025extended}, our method furthermore adopts the log-normal semi-axes representation to ensure only positive axes have support. In contrast, all literature algorithms directly use a Gaussian distribution

As discussed in Section~\ref{sec:introduction}, an alternative modeling approach for \ac{eot} is based on random matrices. Originally proposed in~\cite{koch2008bayesian} and further developed in~\cite{feldmann2010tracking}, the method has seen widespread application. A variety of extensions have been proposed~\cite{lan2017tracking}, e.g., incorporating the measurement number into the shape estimation~\cite{lan2023extended} or incorporating physical constraints~\cite{pei2025constrained}.
One key issue of the \ac{rm} approach is the fact that the entire shape uncertainty is captured by a scalar value. In particular, this means that it is not directly possible to model different process uncertainties for the axis and orientation, as needed for turning objects with fixed size.
A remedy has been proposed in~\cite{tuncer2021random}, where an iterative approach based on variational Bayes with particular focus on estimating the orientation has been proposed. 
Variational Bayes methods for \ac{eot} have also been explored in~\cite{zhang2024extended, wang2025extended}, and in~\cite{zhang2025adaptive}, which focused on group target tracking based on a set of ellipses, rather than a single one, as in standard elliptical \ac{eot}. 
Generally, these methods are not computed in closed-form, and hence are not directly comparable to the proposed approach. 
Notably, the \ac{rm} model inherently only supports positive semi-axes, as their estimates are proportional to the eigenvalues of a symmetric positive definite matrix.

In this article, we focus on the general problem of elliptical \ac{eot}, where the orientation and semi-axis lengths need to be estimated. For specific classes of tracked objects, it can make sense to instead assume correlations between the direction of the velocity and the shape orientation.
Assuming these two fully coincide greatly alleviates the tracking task, which is exploited in, e.g.,~\cite{govaers2019independent, tesori2023lomem}. 
However, it is also possible to make weaker assumptions regarding this, e.g., for drift-like behavior of targets. Such approaches can be found in~\cite{csahin2024random, wen2024velocity, wen2025extended}. 
Another alternative is to assume unknown but unchanging semi-axes, explored in, e.g.,~\cite{li2023tracking}.
We focus on standard assumptions regarding the measurements, such as Gaussian measurement noise. In some applications, these assumptions may not be applicable, e.g., due to heavy-tailed noise. This is treated in~\cite{zheng2025extended} based on an iterative variational adaptation of the \ac{rm} method.

\section{PROBLEM SETTING}~\label{sec:problem_setting}

We are interested in tracking an extended object parameterized by orientation and semi-axis lengths, described at time $k$ by the state
\begin{equation}
    \fullstate_k = \begin{bmatrix} m_{1, k} & m_{2, k} & \dot{m}_{1, k} & \dot{m}_{2, k} & \theta_k & \gamma_{1, k} & \gamma_{2, k} \end{bmatrix}^\tp
    \enspace{,}
    \label{eq:def_full_state}
\end{equation}
consisting of the kinematics $\kinstate_k$ formed jointly by position $\begin{bmatrix} m_{1, k} & m_{2, k}\end{bmatrix}^\tp$ and velocity $\begin{bmatrix}\dot{m}_{1, k} & \dot{m}_{2, k}\end{bmatrix}^\tp$ and the shape.
The shape orientation is given by $\theta_k$.
Importantly, and opposed to related works, the semi-axis lengths ${\semiaxes_k = \begin{bmatrix}l_{1, k} & l_{2, k}\end{bmatrix}^\tp }$ are modeled in logarithmic representation ${\log(\semiaxes_k) = \axisstate_k}$, with the variable $\axisstate_k$ being estimated as part of the state. By this means, the possible values of the semi-axes are restricted to the physically meaningful values in $(0, \inf)$.
Commonly, a representation using two semi-axes and an orientation is used for elliptical targets, but it can also be directly employed for rectangular ones.
We do not make any assumptions on correlations between velocity and orientation as observed for specific target models, and instead treat the general case where they may be entirely independent.
For group object tracking, where multiple objects are tracked as one spatially extended target, this is indispensable, as orientation and velocity are fully detached in that case. 
Furthermore, even for objects which generally are oriented along their course, e.g., cars, there are important corner-cases such as accidents where external force means that such assumptions are invalid. While rare, good tracking accuracy is particularly important in such situations.

In each time step, one or several measurements  
\begin{equation}
    \mat{z}^i_k = \mat{y}^i_k + \boldsymbol{\nu}^i_k   
    \label{eq:meas_based_on_source}
\end{equation} 
are observed. Here, the $i$-th noise-free measurement source $\mat{y}^i_k$ at time $k$ is uniformly distributed across the object extent and $\boldsymbol{\nu}^i_k$ is zero-mean additive noise with $\boldsymbol{\nu}^i_k\sim\mathcal{N}(0, \meascov)$. We assume i.i.d. measurement noise, but all computations can be directly adapted to account for varying measurement noise between time steps or even between individual measurements of a single time step.
The set of $M_k$ measurements at time $k$ is then given by $\mat{Z}_k = \{\mat{z}^i_k~|~i=1\dots M_k\}$. In certain settings, $M_k = 1~\forall k\in\mathbb{N}$, i.e., measurements must be processed sequentially. In the following, this will be referred to as processing a \textit{stream} of measurements.

This measurement setting can also be modeled using the \ac{mem}, which will be briefly reviewed in the following. The \ac{mem} gives rise to the measurement equation~\cite{baum2012modeling}
\begin{equation}
    \mat{z}^i_k = \begin{bmatrix}m_{1, k} \\ m_{2, k}\end{bmatrix} + \rot{\theta_k}\diag\left(l_{1,k}, l_{2,k}\right) \begin{bmatrix}h_1 \\ h_2\end{bmatrix}+ \boldsymbol{\nu}^i_k    
    \enspace,
    \label{eq:mem:def}
\end{equation}
i.e., the measurement source $\mat{y}^i_k$ is modeled using a multiplicative term $\begin{bmatrix}h_1 & h_2\end{bmatrix}^\tp$. Intuitively, the term describes the relative position of the measurement source along the semi-axes of the object, which is then added to the object center position, here written explicitly without a measurement matrix.
Note that this means that the measurement model depends on the logarithm of the state variable $\axisstate$ and that this notation is equivalent to \eqref{eq:meas_based_on_source}. The multiplicative term in~\eqref{eq:mem:def} merely models the measurement source, but the measurement noise is still additive zero-mean Gaussian noise.
The distribution of $h_1$ and $h_2$ is crucial for modeling the distribution of measurement sources on the object surface. For the previously discussed setting of uniform measurement across the full ellipse,
\begin{equation}
    \begin{bmatrix}h_1 \\ h_2\end{bmatrix} \sim \mathcal{N}\left(0, \scalingfactor \cdot \mat{I}\right)
    \enspace,
    \label{eq:mem:dist}
\end{equation}
with scaling factor $\scalingfactor = 0.25$~\cite{yang2019tracking}. For a uniform distribution on a rectangle, we get $\scalingfactor = \frac{1}{3}$ instead. We refer to~\cite{steuernagel2024random} for a detailed discussion on scaling factors.

The state evolves according to a discrete state transition model. For brevity, in this article, we focus on linear motion models, e.g., the constant velocity model. As discussed below, the incorporation of nonlinear motion models is directly possible.

\section{METHODOLOGY}~\label{sec:methodology}
Estimating the full state from~\eqref{eq:def_full_state} jointly requires a number of approximations~\cite{yang2019tracking} and is commonly approached with iterative methods, e.g., variational Bayes~\cite{tuncer2021random}. 
An alternative is to decouple the state into separate components.
Given a prior estimate for the individual parts, fewer approximations are needed, and more accurate individual estimators can be derived.
Depending on the application, a modular algorithm can furthermore be advantageous w.r.t. the incorporation of alternate models, if only individual parts need to be exchanged.
For example, if the shape and velocity orientation of the object are known to coincide, the orientation estimation may be foregone entirely, replacing it with the orientation estimate from the kinematic estimator. For Cartesian velocity models, this can be achieved using the $\arctan$ of the velocity vector. If motion models that explicitly incorporate the heading are used, e.g., the coordinated turn model, the orientation can be directly used.

Therefore, we decouple the full state into kinematics, orientation, and size and provide an estimator for each of the respective parts.
All three components are modeled as independent random variables represented by their first two moments.
Importantly, in literature~\cite{yang2019tracking, tesori2023lomem, steuernagel2025extended}, the semi-axes are directly modeled as Gaussians, therefore implying support for semi-axes in $(-\inf, \inf)$, despite non-positive semi-axes having no physical meaning.
Here, we model $\axisstate_k$ as Gaussian, with $\semiaxes_k = \exp(\axisstate_k)$. This results in a log-normal semi-axes distribution, which only has support for axis in $(0, \inf)$, i.e., physically meaningful size estimates.
The three estimates at time $k$ for the kinematics, semi-axes in log-normal form and orientation are given by $\hat{\kinstate}_k, \hat{\axisstate}_k$ and $\hat{\theta}_k$, respectively. The corresponding covariances\footnote{Note that $\generalcov^{\theta}_k$, the variance of $\hat{\theta}_k$, is scalar, but for reasons of coherence written in the same form as the multivariate covariance matrices.} are $\generalcov^{\kinstate}_k, \generalcov^{\axisstate}_k$ and $\generalcov^{\theta}_k$.
For the update, we propose two variants of the algorithm named \ac{memeqkf}.

In this section, we follow the sequential approach taken by the original \ac{memekf}~\cite{yang2019tracking}, processing each measurement $\mat{z}^i_k \in \mat{Z}_k$ in succession. The updates are carried out in an iterative fashion, i.e., all three estimates are updated with $\mat{z}^i_k$ before $\mat{z}^{i+1}_k$ is processed.
This means that the order in which the three components are updated with each new measurement does not matter. The update of each component is carried out given the estimates acquired after updating all components with the previous measurement, or with the predicted states for the first measurement.

\begin{figure*}
    \centering
    \includegraphics[width=\textwidth]{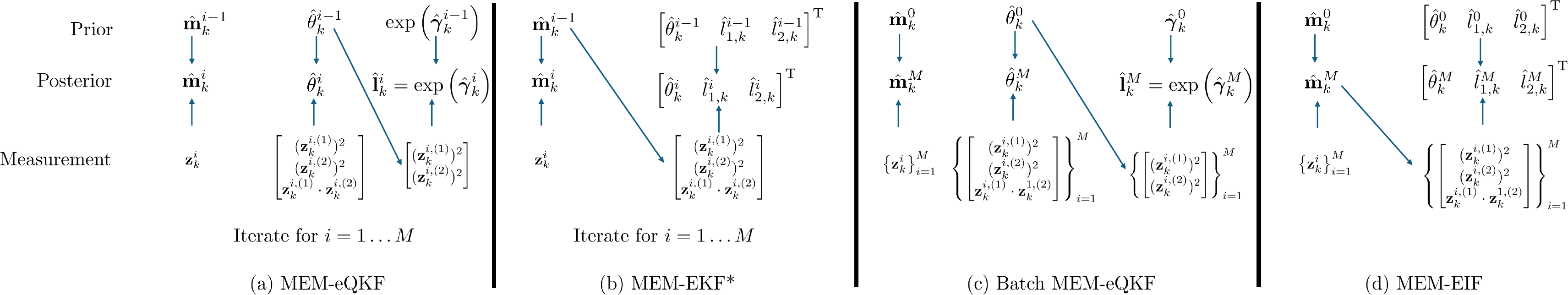}
    \caption{
    Comparison of the proposed and closely related state-of-the-art algorithms~\cite{yang2019tracking, gramsch2024batch}. For each algorithm, the individually updated components of the state are shown along with the (pseudo-)measurements used for the measurement update.
    Arrows indicate input used for individual steps of the computation.
    For the \acrshort{memeqkf}, the variant using~\eqref{eq:meascenter:batch} is shown. The initial prior before processing any measurements is given by $(\cdot)^0$ and the overall posterior by $(\cdot)^M$.
    }
    \label{fig:algo-overview}
\end{figure*}

Afterwards, in Section~\ref{sec:batch}, we will discuss a batch-based update, where all measurements $\mat{Z}_k$ are incorporated in a single correction step. This approach follows the pattern proposed by the \ac{memeif}~\cite{gramsch2024batch} and results in the Batch \ac{memeqkf}.
A summary of the two variants of the proposed algorithm and their related reference algorithms is given in Fig.~\ref{fig:algo-overview}. Here, the measurement update steps are shown schematically, along with the \mbox{(pseudo-)}measurements used during the update. 
The proposed decoupled approach compared to the previous state-of-the-art is clearly visible. This allows avoiding the need for incorporation of the cross-term for the axis update, giving rise to a more accurate estimator.

In the following, a discussion of the log-normal state representation will be given in Section~\ref{sec:method:prelim-axis}. Afterwards, the predict and update steps for each of the three components will be described in Sections~\ref{sec:method:kinematics}, \ref{sec:method:axis}, and \ref{sec:method:orientation}. In between, the necessary pseudo-measurements for the quadratic filters used in the shape update will be introduced in Section~\ref{sec:method:shape_measurements}.
A summary of the algorithm is given in Algorithm~\ref{alg:sequential}.

\begin{algorithm}[t]
\caption{
\mbox{Sequential \acrshort{memeqkf}}
}
\label{alg:sequential}
\begin{enumerate}
    \item \textbf{Predict:} Compute $\hat{\kinstate}^{0}_k$, $\generalcov^{\kinstate, 0}_k$, $\hat{\axisstate}^{0}_k$, $\generalcov^{\axisstate, 0}_{k}$, $\hat{\theta}^{0}_k$, $\generalcov^{\theta, 0}_{k}$ using three standard predict steps following \eqref{eq:kinematic:predict}
    with the corresponding process noise covariance matrices for each component.
    \item \textbf{Update:} For each $\mat{z}^i_k\in \mat{Z}_k$, $i=1\dots|\mat{Z}_k|$:
    \begin{enumerate}
        \item[\textbullet] \textit{For the update, the prior of each component \\ is given by the posterior after processing measurement $\mat{z}^{i-1}_k$},
        \item Compute kinematics $\hat{\kinstate}^i_k, \generalcov^{\kinstate, i}_k$ using \eqref{eq:kinematic:update},
        \item Compute zero-centered measurements $\mat{s}^i_k$ \\ using \eqref{eq:meascenter:batch} (or \eqref{eq:meascenter:stream} for $|\mat{Z}_k|=1$), and pseudo-measurements using \eqref{eq:pseudomeas:axis} and \eqref{eq:pseudomeas:orientation},
        \item Compute logarithmic semi-axis lengths $\hat{\axisstate}^i_k$, \\$ \generalcov^{\axisstate, i}_{k}$ using \eqref{eq:axis:update},
        \item Compute orientation $\hat{\theta}^i_k, \generalcov^{\theta, i}_{k}$ using \eqref{eq:theta:update},
    \end{enumerate}
    \item The posterior is given by the results of the last step, i.e., $\hat{\kinstate}^{|\mat{Z}_k|}_k$, $\generalcov^{\kinstate, |\mat{Z}_k|}_k$ for the kinematics,\\ $\hat{\axisstate}^{|\mat{Z}_k|}_k$, $\generalcov^{\axisstate, |\mat{Z}_k|}_{k}$ for the (logarithmic) axis lengths, \\and $\hat{\theta}^{|\mat{Z}_k|}_k$, $\generalcov^{\theta, |\mat{Z}_k|}_{k}$ for the orientation.
    \item If needed, acquire semi-axes from~\eqref{eq:staterep:gamma-to-axis}
\end{enumerate}
\end{algorithm}

\subsection{Semi-axes representation}\label{sec:method:prelim-axis}
In general, an elliptical extended object tracker is expected to produce estimates of the semi-axes $\semiaxes$, instead of $\axisstate$. Given the estimates tracked in the proposed framework, the semi-axes can directly be recovered using standard formulas for the log-normal distribution. 
Given $\gamma\sim\mathcal{N}(\hat{\gamma}, \sigma^2_{\gamma})$ with the semi-axis $l = \exp(\gamma)$, the mean and variance of the semi-axis distribution can be acquired as
\begin{subequations}
\begin{align}
    \begin{split}
        \hat{l} = \exp\left(\hat{\gamma} + \frac{1}{2}\sigma^2_{\gamma}\right)
    \end{split}        
    \\*
    \begin{split}
        \sigma^2_l = \left(\exp(\sigma^2_{\gamma})-1\right)\cdot\hat{l}^2
    \end{split}
\end{align}
\label{eq:staterep:gamma-to-axis}%
\end{subequations}
It can be of interest to incorporate a (Gaussian) prior for the semi-axes into the proposed log-normal filter, e.g., for compatibility with existing models. 
Given a semi-axis length $\hat{l}$ with variance $\sigma^2_l$, the log-normal form is
\begin{subequations}
\begin{align}
    \begin{split}
        \sigma^2_{\gamma} = \log\left(1+\frac{\sigma^2_l}{\hat{l}^2}\right)
    \end{split}
    \\*
    \begin{split}
        \hat{\gamma} = \log(\hat{l}) - \frac{1}{2}\cdot\sigma^2_{\gamma}
    \end{split}   
\end{align}
\label{eq:staterep:axis-to_gamma}%
\end{subequations}
These follow directly as the inverse of~\eqref{eq:staterep:gamma-to-axis}.

Following the overall decoupled paradigm, for simplicity, we here make use of the univariate formulations, applying them separately to each semi-axis. 
In all models used in the experiments in this article, no correlations between semi-axes are induced.
Nevertheless, we provide the filtering formulas already in matrix form. If correlations between semi-axis should be incorporated, e.g., via the process model, one can simply make use of the standard multi-variate equivalents of~\eqref{eq:staterep:gamma-to-axis} and~\eqref{eq:staterep:axis-to_gamma} for the log-normal distribution.
\subsection{Kinematics}\label{sec:method:kinematics}
As described above, $\hat{\kinstate}^0_{k}$ and $\generalcov^{\kinstate, 0}_{k}$ are given by the predicted estimate $\hat{\kinstate}_{k|k-1}$ and $\kincov_{k|k-1}$, acquired using the standard Kalman filter prediction step
\begin{subequations}
\label{eq:kinematic:predict}
\begin{align}
    \begin{split}
        \hat{\kinstate}_{k|k-1} = \mat{F}^{\kinstate}\hat{\kinstate}_{k-1|k-1}
     \enspace,
    \end{split}
    \\
    \begin{split}
        \kincov_{k|k-1} = \mat{F}^{\kinstate}\kincov_{k-1|k-1}\left(\mat{F}^{\kinstate}\right)^\tp + \mat{Q}^{\kinstate}
        \enspace,
    \end{split}
\end{align}
\end{subequations}
with $\mat{Q}^{\kinstate}$ the covariance matrix of the additive process noise and $\mat{F}^{\kinstate}$ the kinematic state transition matrix.

The kinematic update can be directly performed using the standard Kalman filter update. Given the measurement equation discussed in Section~\ref{sec:problem_setting}, the distribution of an individual measurement $\mat{z}^i_k$ is composed of the distribution of the measurement source on the object and the additive noise.
The update of the kinematic components makes use of the measurement matrix
\begin{equation}
    \mat{H} = \begin{bmatrix}
        1 & 0 & 0 & 0 \\ 0 & 1 & 0 & 0
    \end{bmatrix}
    \enspace,
\end{equation}
which selects the object center from the four-dimensional kinematic state vector and is given by
\begin{subequations}
\label{eq:kinematic:update}
\begin{align}
    \begin{split}
        \hat{\kinstate}^i_k &= \hat{\kinstate}^{i-1}_k + \mat{K}^i_k \left(\mat{z}^i_k-\mat{H}\hat{\kinstate}^{i-1}_k\right)
        \enspace,
    \end{split}
    \\
    \begin{split}
        \generalcov^{\kinstate, i}_k &= \generalcov^{\kinstate, i-1}_k - \mat{K}^i_k\mat{H}\generalcov^{\kinstate, i-1}_k
        \enspace,
    \end{split}
    \\
    \begin{split}
        \mat{K}^i_k &= \generalcov^{\kinstate, i-1}_k \mat{H}^\tp \left(\mat{H}\generalcov^{\kinstate, i-1}_k\mat{H}^\tp + \tilde{\mat{R}}_k\right)^{-1}
        \enspace.
    \end{split}
\end{align}
\end{subequations}
For the measurement noise $\tilde{\mat{R}}_k$, both the additive Gaussian noise as well as the shape of the target must be accounted for. 
This treats the (zero-mean) multiplicative term from \eqref{eq:mem:def}.
As discussed in~\cite{feldmann2010tracking}, this is facilitated by setting
\begin{equation}
    \tilde{\mat{R}}_k = \meascov + \scalingfactor\mat{X}_k
    \enspace,
    \label{eq:kinematic_noise_with_shape}
\end{equation}
with the shape matrix $\mat{X}_k \in \mathbb{R}^{2\times2}$ representing the object extent at time $k$ and scaling factor $\scalingfactor$. For uniform measurements from an elliptic target, $\scalingfactor=0.25$~\cite{feldmann2012comments}, corresponding to the variance of the multiplicative term in \eqref{eq:mem:dist}. 
As $\mat{X}_k$ is not known, we instead use the current estimate $\hat{\mat{X}}^{i-1}_k$ like~\cite{feldmann2010tracking}, given by
\begin{subequations}
\begin{align}
    \begin{split}
    \hat{\mat{X}}^{i-1}_k &= \rot{\hat{\theta}^{i-1}_k} 
    \hat{\mat{D}}^{i-1}_k
    \rot{\hat{\theta}^{i-1}_k}^\tp
     \enspace,
     \label{eq:est_shape_mat}
    \end{split}
    \\
    \begin{split}
        \hat{\mat{D}}^{i-1}_k &=
        \diag\left(
        \left(\hat{\semiaxes}^{{i-1}, (1)}_k\right)^2,
        \left(\hat{\semiaxes}^{{i-1}, (2)}_k\right)^2
        \right)
        \label{eq:def_prior_axis_diag}
        \enspace.
    \end{split}
\end{align}
\end{subequations}
Importantly,~\eqref{eq:def_prior_axis_diag} requires the semi-axes $\semiaxes_k$, not $\axisstate_k$, i.e., conversion using~\eqref{eq:staterep:gamma-to-axis} is needed.
The result after computing the final measurement corresponds to the posterior estimate, i.e., $\hat{\kinstate}_{k|k} = \hat{\kinstate}^{M_k}_k$ and equivalently for the covariance.

\subsection{Pseudo-Measurements}\label{sec:method:shape_measurements}
For the estimation of orientation and size, measurements are zero-centered. 
This is common in literature: For example, the well-established \ac{rm}~\cite{feldmann2010tracking} approach updates the shape using a pseudo-measurement computed using the sample covariance of measurements. Of course, this is computed from the difference of measurements to their mean, with the mean representing an estimate for the object center.
In~\cite{gramsch2024batch}, the posterior of the object center is subtracted from the measurements.
In the following, we will use $\mat{s}^i_k$ as the zero-centered measurements for the shape update. 
If more than one measurement is available at each time step, these can simply be computed as
\begin{equation}
    \mat{s}^{i}_k = \mat{z}^i_k - \bar{\mat{z}}_k \qquad i=1\dots M_k
    \label{eq:meascenter:batch}
    \enspace,
\end{equation}
with $\bar{\mat{z}}_k$ being the mean of measurements at time $k$. 
If, however, only a single measurement is received,~\eqref{eq:meascenter:batch} would always yield zero. 
As an alternative, the prior estimate of the object center $\mat{H}\hat{\kinstate}_{k|k-1}$ can be used instead of the mean of measurements, giving rise to
\begin{equation}
        \mat{s}^{i, \textit{stream}}_k = \mat{z}^i_k - \mat{H}\hat{\kinstate}_{k|k-1} \qquad i=1
    \label{eq:meascenter:stream}
    \enspace.
\end{equation}
This additionally introduces the uncertainty of the predicted center into the pseudo-measurement, accounted for by using
\begin{equation}
    \mat{W}_k = \begin{cases}
        \mat{R},& \text{if } M_k > 1\\
        \mat{R}+\mat{H}\kincov_{k|k-1}\mat{H}^\tp,& \text{otherwise.}
    \end{cases}
    \label{eq:centered_uncertainty}
\end{equation}
As originally proposed in~\cite{baum2012modeling} and further exploited in~\cite{yang2019tracking, steuernagel2025extended}, we make use of quadratic pseudo-measurements for the shape update. For the axis, it is sufficient to employ
\begin{subequations}
\begin{align}
    \begin{split}
        \mat{a}^i_k &= \mat{a}^{i, \text{sqrt}}_k \odot \mat{a}^{i, \text{sqrt}}_k
        \enspace,
    \end{split}
    \\*
    \begin{split}
        \mat{a}^{i, \text{sqrt}}_k &= 
        \rot{-\theta^{i-1}_k}
        \begin{bmatrix}
            \left(\mat{s}^{i, (0)}_k\right) &
            \left(\mat{s}^{i, (1)}_k\right)
        \end{bmatrix}^\tp
        \enspace,
    \end{split}
\end{align}
\label{eq:pseudomeas:axis}%
\end{subequations}
where $\odot$ is the element-wise product, i.e., the centered measurements $\mat{s}$ are rotated based on prior orientation estimate $-\theta^{i-1}_k$ and then squared element-wise.
For the orientation, the cross-term is incorporated as 
\begin{equation}
    \renewcommand*{\arraystretch}{1.5}
    \mat{b}^i_k = 
    \begin{bmatrix}
        \left(\mat{s}^{i, (0)}_k\right)^2 &
        \left(\mat{s}^{i, (1)}_k\right)^2 &
        \mat{s}^{i, (0)}_k \cdot \mat{s}^{i, (1)}_k
    \end{bmatrix}^\tp
    \enspace.
    \label{eq:pseudomeas:orientation}
\end{equation}

\subsection{Axis Estimation}\label{sec:method:axis}
If the state transition model is directly available for $\axisstate$, i.e., in logarithmic representation, a standard Kalman filter predict step can be directly carried out.
When models designed for reference algorithms are used, both the state transition and the process noise covariance matrix for the semi-axes might, however, be given directly the axis-domain, rather than the logarithmic representation.
Using~\eqref{eq:staterep:gamma-to-axis} and~\eqref{eq:staterep:axis-to_gamma}, the current state estimate and its covariance can be converted to the semi-axes representation, where a predict step using equivalent to~\eqref{eq:kinematic:predict}, but using $\mat{F}^{\axisstate}$ and $\mat{Q}^{\axisstate}$, can be carried out.
For standard models as considered here, ${\mat{F}^{\axisstate} = \mat{I}^{2\times 2}}$, the two-by-two identity matrix. Non-zero values of $\mat{Q}^{\axisstate}$ can account for the object size changing over time.
For comparison with the reference algorithms representing the semi-axes directly, we here adopt the conversion-based approach.

The update equations are given by~\cite{baum2012modeling, steuernagel2025extended}
    \begin{subequations}
    \begin{align}
    \begin{split}
        \hat{\axisstate}^i_k &= \hat{\axisstate}^{i-1}_k 
        + \mat{C}^{\mat{a}, \axisstate}_{k, i}
        \left(\mat{C}^{\mat{a}, \mat{a}}_{k, i}\right)^{-1}
        \left(
        \mat{a}^i_k - \EX(\mat{a}^i_k)
        \right)
        \enspace,
        \label{eq:axis:update:state}
    \end{split}
    \\*
    \begin{split}
        \generalcov^{\axisstate, i}_{k} &= \generalcov^{\axisstate, i-1}_{k} 
        - \mat{C}^{\mat{a}, \axisstate}_{k, i}
        \left(\mat{C}^{\mat{a}, \mat{a}}_{k, i}\right)^{-1}
        \left(\mat{C}^{\mat{a}, \axisstate}_{k, i}\right)^\tp
        \enspace.
        \label{eq:axis:update:cov}
    \end{split}
\end{align}
\label{eq:axis:update}%
\end{subequations}
The key difference to previous works~\cite{steuernagel2025extended, steuernagel2025decoupled} is that, as the axis representation in the state is in log-normal form, the moments $\EX(\mat{a}^i_k), \mat{C}^{\mat{a}, \mat{a}}_{k, i}$ and $\mat{C}^{\mat{a}, \axisstate}_{k, i}$ need to be adapted accordingly. 
For all three, the derivation is presented in Appendix~\ref{app:moments}. 
For brevity, we will make use of 
\begin{equation}
    \EX\left[\exp(\axisstate^{i}_k)^2\right]^{(j)} = \exp\left(2\cdot\hat{\axisstate}^{i, (j)}_{k}+2\cdot\generalcov^{\hat{\axisstate}, i, (j, j)}_{k}\right)
    \label{eq:memkf:expected-square}
    \enspace,
\end{equation}
which is the j-th entry of the expected square of the log-normal distribution representing the axis state at time $k$ and iteration $i$.
In summary, the required matrices and expected pseudo-measurement are given by
\begin{equation}
        \renewcommand*{\arraystretch}{2}
        \EX(\mat{a}^i_k) = \begin{bmatrix}
         \scalingfactor\cdot
         \EX\left[\exp(\axisstate^{i}_k)^2\right]^{(1)} 
         + \mat{W}^{\theta, (1, 1)}_k \\
        \scalingfactor\cdot
         \EX\left[\exp(\axisstate^{i}_k)^2\right]^{(2)} 
         + \mat{W}^{\theta, (2, 2)}_k 
        \end{bmatrix}
        \enspace.
        \label{eq:memkf:exp}
\end{equation}
The off-diagonal elements of the pseudo-measurement covariance matrix are
\begin{subequations}
    \begin{align}
        \begin{split}
            \left(\mat{C}^{\mat{a}, \mat{a}}_{k, i}\right)^{(1, 2)} &= 2\cdot\left(\mat{W}^{\theta, (1, 2)}_k\right)^2
        \end{split}
        \\
        \begin{split}
            \left(\mat{C}^{\mat{a}, \mat{a}}_{k, i}\right)^{(2, 1)} &= 2\cdot\left(\mat{W}^{\theta, (2, 1)}_k\right)^2
        \end{split}
    \end{align}
\end{subequations}
where 
\begin{equation}
    \mat{W}^{\theta}_k = \rot{-\hat{\theta}^{i-1}_k}~\mat{W}_k~\rot{-\hat{\theta}^{i-1}_k}^\tp
\end{equation}
is the axis-aligned version of the measurement covariance $\mat{W}_k$ from~\eqref{eq:centered_uncertainty}. 
The $j$-th diagonal element with $j=1, 2$ is given by
\begin{align}
    \begin{split}
        &\left(\mat{C}^{\mat{a}, \mat{a}}_{k, i}\right)^{(j, j)}
        = 3\cdot\scalingfactor^2\cdot\exp\left(4\cdot\hat{\gamma}_j + 8\cdot(\generalcov^{\axisstate, i}_{k})^{(j, j)}\right)
    \end{split} \nonumber
    \\
    \begin{split}
        &\quad+ 6\cdot\scalingfactor\cdot\exp\left(2\cdot\hat{\gamma}_j + 2\cdot(\generalcov^{\axisstate, i}_{k})^{(j, j)}\right)\cdot\mat{W}^{\theta, (j, j)}
    \end{split}\nonumber
        \\
    \begin{split}
        &\quad+ 3\cdot\left(\mat{W}^{\theta, (j, j)}\right)^2- \left(\EX(\mat{a})^{(j)}\right)^2
        \enspace.
    \end{split}
\end{align}
The cross-covariance between $\mat{a}_k$ and $\axisstate_k$ given by $\mat{C}^{\mat{a}, \axisstate}_{k, i}$ is a diagonal matrix with entries
\begin{subequations}
\begin{align}
    \begin{split}
        \left(\mat{C}^{\mat{a}, \axisstate}_{k, i}\right)^{(1, 1)} &= 
        2\cdot\scalingfactor\cdot
        \generalcov^{\hat{\axisstate}, i-1, (1, 1)}_{k}\cdot
        \EX\left[\exp(\axisstate^{i}_k)^2\right]^{(1)}
        \enspace,
    \end{split}
    \\*
    \begin{split}
       \left(\mat{C}^{\mat{a}, \axisstate}_{k, i}\right)^{(2, 2)} &=         2\cdot\scalingfactor\cdot
        \generalcov^{\hat{\axisstate}, i-1, (2, 2)}_{k}\cdot
        \EX\left[\exp(\axisstate^{i}_k)^2\right]^{(2)}
        \enspace.
    \end{split}
\end{align}
\label{eq:axis:crosscovmatrix}%
\end{subequations}

\subsection{Orientation Estimation}\label{sec:method:orientation}
As before, the prediction of the orientation is a standard prediction step for the one-dimensional estimate, computing $\hat{\theta}_{k|k-1}$, $\generalcov^{\theta}_{k|k-1}$.

To compute the updated orientation estimate $\hat{\theta}_{k|k}$ and its variance $\generalcov^{\theta}_{k|k}$, we make use of (parts of) the results presented in~\cite{yang2019tracking}, where the \ac{memekf} was originally described. The update is carried out analogously to Section~\ref{sec:method:axis}, but for the second pseudo-measurement $\mat{b}^i_k$ and correspondingly different realizations for the expected pseudo-measurement and the two covariances. 
It reads
\begin{subequations}
\begin{align}
    \begin{split}
        \hat{\theta}^i_k &= \hat{\theta}^{i-1}_k 
        + \mat{C}^{\mat{b}, \theta}_{k, i}
        \left(\mat{C}^{\mat{b}, \mat{b}}_{k, i}\right)^{-1}
        \left(
        \mat{b}^i_k - \EX(\mat{b}^i_k)
        \right)
        \enspace,
        \label{eq:theta:update:state}
    \end{split}
    \\*
    \begin{split}
        \generalcov^{\theta, i}_{k} &= \generalcov^{\theta, i-1}_{k} 
        - \mat{C}^{\mat{b}, \theta}_{k, i}
        \left(\mat{C}^{\mat{b}, \mat{b}}_{k, i}\right)^{-1}
        \left(\mat{C}^{\mat{b}, \theta}_{k, i}\right)^\tp
        \enspace.
        \label{eq:theta:update:cov}
    \end{split}
\end{align}
\label{eq:theta:update}%
\end{subequations}
Computation of the three required expectations is slightly more involved, as the derivation requires linearization. For brevity, we will omit detailed explanations and refer to~\cite{yang2019tracking} for a comprehensive discussion and derivation.
As with~\eqref{eq:def_prior_axis_diag}, the orientation update requires the semi-axes $\semiaxes_k$, not the log-normal form $\axisstate_k$, meaning~\eqref{eq:staterep:gamma-to-axis} needs to be used.

The definition of the expectations makes use of
\begin{subequations}
\label{eq:theta:helper1}
\begin{align}
    \begin{split}
       \mat{V} &= \begin{bmatrix}
           1 & 0 & 0 & 0 \\
           0 & 0 & 0 & 1 \\
           0 & 1 & 0 & 0 \\
       \end{bmatrix}
       ~,~
     \tilde{\mat{V}} = \begin{bmatrix}
           1 & 0 & 0 & 0 \\
           0 & 0 & 0 & 1 \\
           0 & 0 & 1 & 0 \\
       \end{bmatrix}
       \enspace,
    \end{split}
    \\*
    \begin{split}
        \mat{C}_h &= \scalingfactor\cdot\mat{I}^{2\times 2}
        \enspace,
    \end{split}
    \\*
    \begin{split}
        \mat{S}^i_k &= \rot{\hat{\theta}^{i-1}_k} \hat{\mat{D}}^{i-1}_k
        \enspace,
    \end{split}
    \\*
    \begin{split}
        \mat{J}^i_{k, 1} &= \begin{bmatrix}
            -\hat{\semiaxes}^{i-1, (1)}_k \cdot \sin(\hat{\theta}^{i-1}_k) &
            -\hat{\semiaxes}^{i-1, (2)}_k \cdot \cos(\hat{\theta}^{i-1}_k)
        \end{bmatrix}^\tp
        \enspace,
    \end{split}
    \\*
    \begin{split}
        \mat{J}^i_{k, 2} &= \begin{bmatrix}
            \hat{\semiaxes}^{i-1, (1)}_k \cdot \cos(\hat{\theta}^{i-1}_k) &
            -\hat{\semiaxes}^{i-1, (2)}_k \cdot \sin(\hat{\theta}^{i-1}_k)
        \end{bmatrix}^\tp
        \enspace,
    \end{split}
\end{align}
\end{subequations}
with $\hat{\mat{D}}^{i-1}_k$ as given in~\eqref{eq:def_prior_axis_diag}.
Furthermore, $\mat{S}^i_k$ is divided in its first and second row, in the following denoted by $\mat{S}^i_{k,1}$ and $\mat{S}^i_{k,2}$, respectively.
Based on~\eqref{eq:theta:helper1}, we moreover define
\begin{subequations}
\label{eq:theta:helper2}
\begin{align}
    \begin{split}
       \mat{C}^{i}_{k, \text{I}} &= \mat{S}^i_k \mat{C}_h \left(\mat{S}^i_k\right)^\tp
       \enspace,
    \end{split}
    \\*
    \begin{split}
        \mat{C}^{i, (m,n)}_{k, \text{II}} 
        &= 
        \trace\left(\generalcov^{\theta, i-1}_{k} \left(\mat{J}^i_{k, n}\right)^\tp \mat{C}_h \mat{J}^i_{k, m}\right),~ m, n \in \{1, 2\}
        \enspace,
    \end{split}
    \\*
    \begin{split}
        \mat{C}^{i}_{k, \mat{s}} &= \mat{W}_k + \mat{C}^{i}_{k, \text{I}} + \mat{C}^{i}_{k, \text{II}}
        \enspace,
        \label{eq:orientation:cov_centered_meas}
    \end{split}
    \\*
    \begin{split}
        \mat{M}^{i}_{k} &= \begin{bmatrix}
            2 \cdot \mat{S}^i_{k,1} \mat{C}_h \mat{J}^i_{k, 1}  \\
            2 \cdot \mat{S}^i_{k,2} \mat{C}_h \mat{J}^i_{k, 2} \\
            \mat{S}^i_{k,1} \mat{C}_h \mat{J}^i_{k, 2} + \mat{S}^i_{k,2} \mat{C}_h \mat{J}^i_{k, 1}
        \end{bmatrix}
        \enspace.
    \end{split}
\end{align}
\end{subequations}
Finally, the three desired expectations are given by
\begin{subequations}
\begin{align}
    \begin{split}
       \EX(\mat{b}^i_k) &= \mat{V} \vect\left(\mat{C}^{i}_{k, \mat{s}}\right)
       \label{eq:orientation:expected_pseudomeas}
    \end{split}
    \\*
    \begin{split}
        \mat{C}^{\mat{b}, \mat{b}}_{k, i} &= \mat{V} \left(\mat{C}^{i}_{k, \mat{s}}\otimes\mat{C}^{i}_{k, \mat{s}}\right) (\mat{V}+\tilde{\mat{V}})^\tp
        \enspace,
        \label{eq:orientation:pseudomeascov}
    \end{split}
    \\*
    \begin{split}
        \mat{C}^{\mat{b}, \theta}_{k, i} &= \generalcov^{\theta, i-1}_{k} \mat{M}^{i}_{k}
        \enspace,
    \end{split}
\end{align}
\label{eq:orientation:expectations}%
\end{subequations}
where $\vect$ creates a column vector from a matrix by stacking its columns as defined in~\cite{yang2019tracking} and $\otimes$ is the Kronecker product.

\section{BATCH-BASED FORMULATION}\label{sec:batch}
In the previous section, the proposed algorithm was introduced in a sequential manner, i.e., all measurements are explicitly iterated once in order to compute the overall posterior.
This is significantly more efficient than algorithms that perform multiple passes over the set of measurements, such as the iterative \ac{vbrm}~\cite{tuncer2021random} or the sample-based \ac{memrbpf}~\cite{steuernagel2025extended}. 
Nevertheless, for practical applications, even better computational efficiency can be desirable. This can be achieved by performing the measurement update \textit{batch-based}, i.e., performing a single update using all available measurements. 
One of the most prominent algorithms using this strategy is the highly efficient original \ac{rm} algorithm~\cite{feldmann2010tracking}. 
A recent example is the \ac{memeif}~\cite{gramsch2024batch}, which is a batch-based variant of the \ac{memekf}~\cite{yang2019tracking}. 

In the following sections, we will present the batch-based formulation of the proposed \ac{memeqkf}. This variant follows the same approach as the previous section, decoupling the kinematics, orientation and semi-axis lengths. However, each of the three posteriors is computed using a single Kalman filter-based update making use of all measurements. 
Therefore, the iteration index $i$ is omitted, and instead explicit notation of the posterior ($k|k$) and prior ($k|k-1$) is used.

For the case that a single measurement is available, the update from the previous section can readily be employed, avoiding the approximations needed for the derivation of the batch-based update.
For $|\mat{Z}_k| > 1$, the batch-based algorithm is described in the following, and summarized in Algorithm~\ref{alg:batch}.

\begin{algorithm}[t]
\caption{
\mbox{Batch-based \acrshort{memeqkf}}
}
\label{alg:batch}
\begin{enumerate}
    \item \textbf{Predict:} Compute $\hat{\kinstate}^{0}_k$, $\generalcov^{\kinstate, 0}_k$, $\hat{\axisstate}^{0}_k$, $\generalcov^{\axisstate, 0}_{k}$, $\hat{\theta}^{0}_k$, $\generalcov^{\theta, 0}_{k}$ using three standard predict steps following \eqref{eq:kinematic:predict}
    with the corresponding process noise covariance \\ matrices for each component.
    \item \textbf{Update:}
    \begin{enumerate}
        \item[\textbullet] \textit{If only a single measurement is available, \\ employ the update from Algorithm~\ref{alg:sequential}},
        \item Compute the kinematic posterior $\hat{\kinstate}_{k|k},\generalcov^{\kinstate}_{k|k}$ using~\eqref{eq:batch:kinematic:update},
        \item Compute the mean axis pseudo-\\measurement $\tilde{\mat{a}}_k$ using~\eqref{eq:batch:axis-pseudomeas},
        \item Compute the axis posterior $\hat{\axisstate}_{k|k}, \generalcov^{\axisstate}_{k|k}$ as\\defined in~\eqref{eq:batch:axis:update}
        \item Compute the orientation pseudo-\\measurements using~\eqref{eq:batch:pseudomeas:orientation_kronecker},
        \item Compute the orientation posterior $\hat{\theta}_{k|k}, \generalcov^{\theta}_{k|k}$ from~\eqref{eq:batch:orientation:posterior}.
    \end{enumerate}
\end{enumerate}
\end{algorithm}

\subsection{Kinematic Batch Update}
The batch-based update of the kinematics is straightforward. 
As done by, e.g., the \ac{rm}~\cite{feldmann2010tracking} algorithm, it is carried out using a pseudo-measurement $\bar{\mat{z}}_k\in\mathbb{R}^2$, which is given by the mean of the measurements
\begin{equation}
    \bar{\mat{z}}_k = \frac{1}{M_k}\sum_{i=1}^{M_k}\mat{z}^i_k
    \enspace.
    \label{eq:batch:kinematic:mean}
\end{equation}
The update formula therefore is given by
\begin{subequations}
\label{eq:batch:kinematic:update}
\begin{align}
    \begin{split}
        \hat{\kinstate}_{k|k} &= \hat{\kinstate}_{k|k-1} + \mat{K}_k \left(\bar{z}_k-\mat{H}\hat{\kinstate}_{k|k-1}\right)
        \enspace,
    \end{split}
    \\
    \begin{split}
        \generalcov^{\kinstate}_{k|k} &= \generalcov^{\kinstate}_{k|k-1} - \mat{K}_k\mat{H}\generalcov^{\kinstate}_{k|k-1}
        \enspace,
    \end{split}
    \\
    \begin{split}
        \mat{K}_k &= \generalcov^{\kinstate}_{k|k-1} \mat{H}^\tp \left(\mat{H}\generalcov^{\kinstate}_{k|k-1}\mat{H}^\tp + \frac{1}{M_k}\tilde{\mat{R}}_k\right)^{-1}
        \enspace.
    \end{split}
\end{align}
\end{subequations}
As for the sequential update, $\tilde{\mat{R}}_k$ is given by~\eqref{eq:kinematic_noise_with_shape}. 
Implementations of nonlinear or otherwise more complex kinematic models can directly use $\bar{\mat{z}}_k$ with corresponding covariance $\frac{1}{M_k}\tilde{\mat{R}}_k$.

\subsection{Axis Batch Update}
In order to compute the batch-based update of the semi-axis, similarly to the kinematic update, we make use of the average of the pseudo-measurements given in Section~\ref{sec:method:axis}, given by
\begin{equation}
    \tilde{\mat{a}}_k = \frac{1}{M_k}\sum_{i=1}^{M_k} \mat{a}^i_k
    \enspace,
    \label{eq:batch:axis-pseudomeas}
\end{equation}
with $\mat{a}^i_k$ defined as in~\eqref{eq:pseudomeas:axis}. 
The expected pseudo-measurement is then again given by $\EX(\mat{a}^i_k)$ as in~\eqref{eq:memkf:exp}.

We approximate the squared pseudo-measurements as independent and further approximate the overall problem as linear, giving rise to the standard Kalman filter equations as the solution. Discussion of the impact of this simplifying approximation is given in Section~\ref{sec:simulation:ablation}.
For the implementation of the linear Kalman filter, we again make use of the individual pseudo-measurement marginal measurement covariance $\mat{C}^{\mat{a}, \mat{a}}_{k, i}$ defined in for the sequential update and the cross-covariance between the state and each pseudo-measurement $\mat{C}^{\mat{a}, \axisstate}_{k, i}$ as in~\eqref{eq:axis:crosscovmatrix}.

For the measurement update, two further matrices are required: The (linearized) measurement matrix $\mat{H}^b_k$ and the marginal measurement covariance $\mat{C}^{\tilde{\mat{a}}, \tilde{\mat{a}}}_{k, i}$.
For $\mat{H}^b_k$, we make use of 
\begin{equation}
    \mat{H}^b_k \approx \mat{C}^{\mat{a}, \axisstate}_{k, i} \left(\generalcov^{\axisstate}_{k|k-1}\right)^{-1}
    \enspace,
\end{equation}
which would be exact in the truly linear case.
Similarly, given our linearization, we can approximate
\begin{equation}
    \mat{R}^b_k \approx \mat{C}^{\mat{a}, \mat{a}}_{k, i} - \mat{H}^b_k \generalcov^{\axisstate}_{k|k-1} \left(\mat{H}^b_k\right)^\tp
    \enspace,
\end{equation}
which would again be exact in the linear case. As we have averaged $M_k$ measurements that are approximated as independent, their overall covariance reduces by a factor $M_k$. Therefore, we can compute
\begin{equation}
    \mat{C}^{\tilde{\mat{a}}, \tilde{\mat{a}}}_{k, i} = 
    \frac{1}{M_k}  \mat{R}^b_k + \mat{H}^b_k \generalcov^{\axisstate}_{k|k-1} \left(\mat{H}^b_k\right)^\tp
    \enspace,
    \label{eq:batch:axis:innov-cov}
\end{equation}
which is the approximate marginal measurement (or innovation) covariance for the $M_k$ measurements evaluated at the prior and approximated as independent. 
Based on these results, a standard Kalman filter update can be computed as
\begin{subequations}
\begin{align}
    \begin{split}
        \mat{K}^b_k &= \generalcov^{\axisstate}_{k|k-1} (\mat{H}^b_k)^\tp \left(\mat{C}^{\tilde{\mat{a}}, \tilde{\mat{a}}}_{k, i}\right)^{-1}
        \label{eq:batch:axis:update:gain}
    \end{split}
    \\*
    \begin{split}
        \hat{\axisstate}_{k|k} &= \hat{\axisstate}^{0}_k 
        + \mat{K}^b_k \left(
        \tilde{\mat{a}}_k 
        - \EX(\mat{a}^i_k)
        \right)
        \enspace,
        \label{eq:batch:axis:update:state}
    \end{split}
    \\*
    \begin{split}
        \generalcov^{\axisstate}_{k|k} &= \generalcov^{\axisstate}_{k|k-1} 
        - \mat{K}^b_k \mat{H}^b_k \generalcov^{\axisstate}_{k|k-1}
        \enspace.
        \label{eq:batch:axis:update:cov}
    \end{split}
\end{align}
\label{eq:batch:axis:update}%
\end{subequations}
In summary, by approximating the squared pseudo-measurements as independent, it is possible to leverage the very same moments computed for the sequential update by using their mean. Approximating the problem given these as linear results in a straightforward linear Kalman filter update incorporating all measurements simultaneously.

\subsection{Orientation Batch Update}
Finally, for the orientation, in previous related work~\cite{gramsch2024batch} a closed-form batch update of the orientation has already been derived based on the information form of the update.
For the derivation, we refer to~\cite{gramsch2024batch}, and here only present the resulting formulas when applied for the proposed algorithm. 

A key difference is that~\cite{gramsch2024batch} made use of the kinematic posterior for the shape update. To avoid double counting of measurement information, we instead make use of $\bar{\mat{z}}_k$, i.e., the mean of measurements instead, which is consistent with the computation of pseudo-measurements above in~\eqref{eq:pseudomeas:orientation}. This is discussed in more detail in Section~\ref{sec:double-counting}.
We make use of the same helper variables already defined in the sequential orientation update, i.e.,~\eqref{eq:theta:helper1} and~\eqref{eq:theta:helper2} are again used in the following.

The posterior orientation estimate and its variance are computed as
\begin{subequations}
    \label{eq:batch:orientation:posterior}
    \begin{align}
        \begin{split}
            \hat{\theta}_{k|k} &= \generalcov^{\theta}_{k|k}\xi^{\theta}_{k|k}
            \enspace,
        \end{split}\\
        \begin{split}
            \generalcov^{\theta}_{k|k} &= 
            \left(
            \left(\generalcov^{\theta}_{k|k-1}\right)^{-1}
            + |\mat{Z}_k| \cdot \mat{M}^{\tp}_{k} \left(\generalcov^t_k\right)^{-1} \mat{M}_{k}
            \right)^{-1}
            \enspace,
        \end{split}
    \end{align}
\end{subequations}
where 
\begin{subequations}
\begin{align}
    \begin{split}
        \generalcov^t_k &= \mat{C}^{\mat{b}, \mat{b}}_{k} - \mat{M}_{k}  \generalcov^{\theta}_{k|k-1} \mat{M}_{k}^\tp
        \enspace,
    \end{split}\\
    \begin{split}
        \xi^{\theta}_{k|k} &= \xi^{\theta}_{k|k-1} 
        + \mat{M}^{\tp}_{k} \left(\generalcov^t_k\right)^{-1} 
        \Xi_k
        \enspace,
    \end{split}\\
    \begin{split}
        \Xi_k &= \sum_{i=1}^{|\mat{Z}_k|} \mat{b}^i - \EX(\mat{b}_k) + \mat{M}_{k} \cdot \hat{\theta}_{k|k-1}
        \enspace,
        \label{eq:batch:orientation:sumXi}
    \end{split}\\
    \begin{split}
        \xi^{\theta}_{k|k-1} &= 
        \left(\generalcov^{\theta}_{k|k-1}\right)^{-1} \cdot  \hat{\theta}_{k|k-1}
        \enspace.
    \end{split}
\end{align}
\end{subequations}
Analogously to~\eqref{eq:pseudomeas:orientation}, the pseudo-measurements can be computed using Kronecker notation as
\begin{equation}
    \mat{b}^i = \mat{V} \left(\mat{s}^i_k \otimes \mat{s}^i_k\right)
    \label{eq:batch:pseudomeas:orientation_kronecker}
\end{equation}
and the expected pseudo-measurement equivalently to the sequential update, i.e., as defined in~\eqref{eq:orientation:expected_pseudomeas}.
The pseudo-measurement covariance, $\mat{C}^{\mat{b}, \mat{b}}_{k}$ is computed as in~\eqref{eq:orientation:pseudomeascov}, since the individual pseudo-measurements are identical.

\subsection{Avoiding Double Counting}~\label{sec:double-counting}
One consideration in the implementation of the extended object tracking filters is avoiding double counting of measurement information, in particular w.r.t. the object center.
In the Bayesian filtering paradigm, the measurement update is carried out based on the measurement likelihood $p(\mat{Z}_k|\mat{x}_{k})$, i.e., the likelihood of the measurements given the predicted state estimate at time $k$. 
Many filters, including the one proposed here, decouple the shape and kinematic estimation procedures from each other~\cite{feldmann2010tracking, govaers2019independent, gramsch2024batch, steuernagel2025extended}.
Here, the kinematic estimate and therefore the estimate of the object center is already computed using the likelihood $p(\mat{Z}_k|\mat{m}_{k})$. Therefore, when computing the update for the extent, conditioning on $\mat{m}_{k|k}$ would effectively lead to double counting of $\mat{Z}_k$, as for the shape update, the likelihood of $\mat{Z}_k$ would be conditioned on a variable already itself conditioned on $\mat{Z}_k$.
Viewing Fig.~\ref{fig:algo-overview}, one can see that the \ac{memekf} and both variants of the proposed algorithms only make use of prior state estimates in the computation of the pseudo-measurements, whereas the \ac{memeif} is the only algorithm where the kinematic posterior is used in the computation of the pseudo-measurements for the shape update.
Nevertheless, the resulting more accurate estimate of the center may allow numerically improved shape estimation; however, the theoretical downside is to be considered. 
Furthermore, such design introduces an arbitrary forced order of operations for the update of two variables considered independent.

Therefore, throughout this article, we make use of the measurement mean in~\eqref{eq:meascenter:batch} or the predicted center in~\eqref{eq:meascenter:stream}, to avoid double counting of measurement information.
This is also a key difference to~\cite{gramsch2024batch}, which uses the kinematic posterior for the shape update as an approximation. Replacing this with the prediction causes~\cite{gramsch2024batch} to diverge.

Identical considerations apply to the orientation estimate. While the orientation is used in similar manner for the computation of the pseudo-measurement for the semi-axes, we argue that incorporating the posterior orientation estimation introduces double-counting of measurement information. Additionally, this would introduce arbitrary ordering to the update steps. Hence, we make use of the prior orientation. Note that for the sequential update, the prior orientation is continuously updated as measurements are received.
Empirically, improved results can be achieved by simply double-counting measurement information (see also~\cite{gramsch2024batch}). Reasons for this and the consistency of the resulting estimates are to be explored.

\section{SIMULATION STUDY}~\label{sec:simulation}

\begin{figure}[t]
    \centering
    \includegraphics[width=\columnwidth]{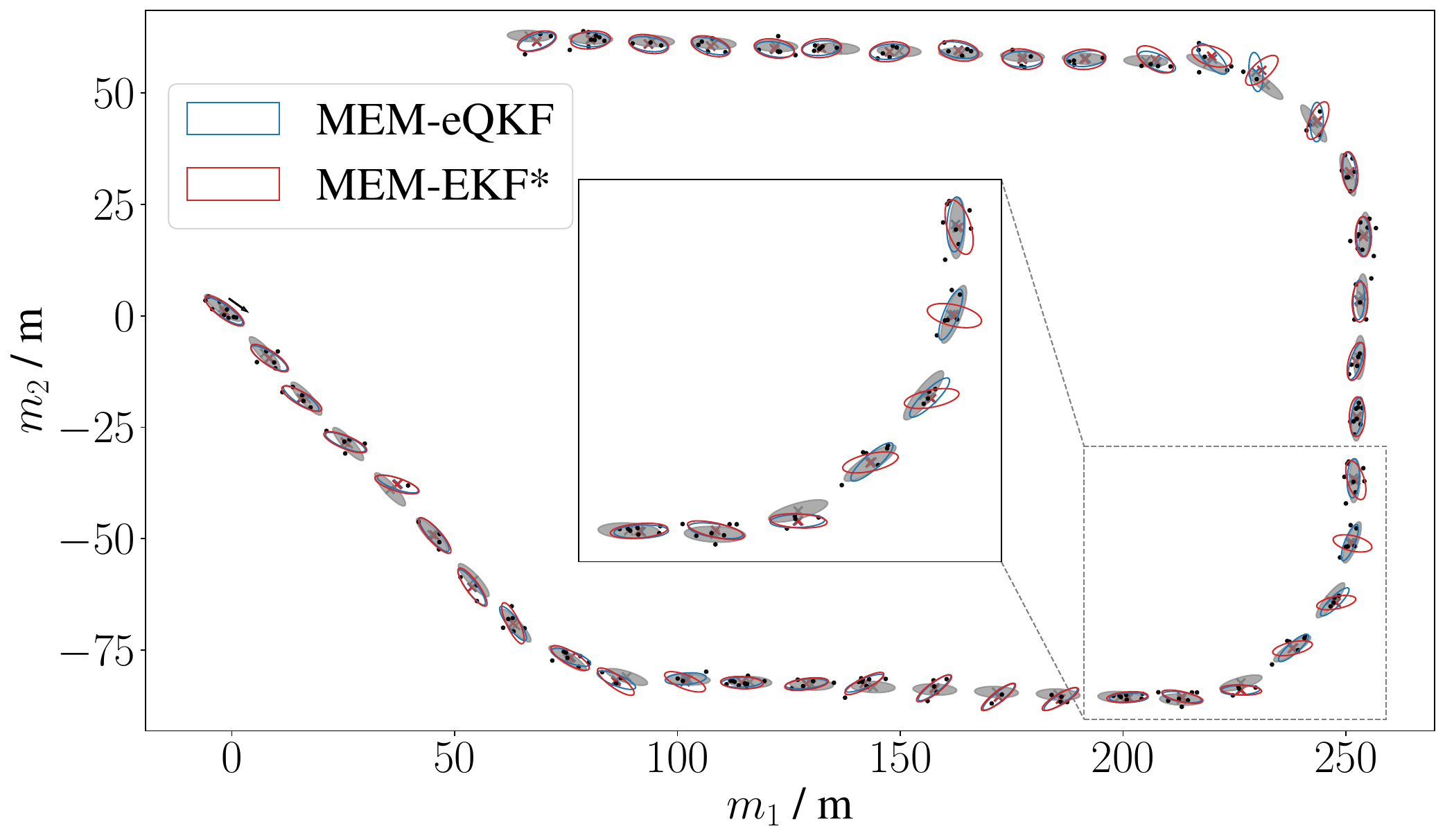}
    \caption{
    Example qualitative results for tracking a maneuvering elliptical object with the proposed \acrshort{memeqkf} and the \acrshort{memekf}~\cite{yang2019tracking} on the simulated trajectory.
    }
    \label{fig:traj_overview}
\end{figure}

To confirm the effectiveness of the proposed method, named \ac{memeqkf}, under the given assumptions, an extensive simulation study was carried out. Two scenarios were assessed: Firstly, a standard simulation based on a target following a pre-defined trajectory (with added zero-mean process noise) as shown in Fig.~\ref{fig:traj_overview}, where measurement parameters were varied to evaluate the trackers' robustness.
Secondly, a stationary setting in which only a single measurement was received in each step, requiring sequential processing of the data.

Multiple existing methods have been chosen as reference algorithms:
The \ac{memekf}~\cite{yang2019tracking}, which is a closed-form algorithm that can be seen as the predecessor of the proposed filter. Furthermore, the closely related \ac{memrbpf}~\cite{steuernagel2025extended}, which is a sampling-based filter from the family of \ac{mem} algorithms.
Additionally, the \ac{vbrm}~\cite{tuncer2021random} has been selected as its explicit focus on orientation estimation makes it a good match for the evaluation.

Of course, the batch-based variant needs to be compared with other batch-based trackers. 
Here, the \ac{rm} tracker~\cite{feldmann2010tracking} has been employed as a baseline algorithm.
Furthermore, the \ac{memeif}~\cite{gramsch2024batch} is a recent state-of-the-art batch-based tracker closely related to the \ac{memekf}\footnote{We have used the \ac{memeif} as-is without further modifications, as removing the heuristic discussed above that leads to double counting of measurement information causes the filter to diverge.}.

To assess the quality of the tracking results jointly for the position and shape estimates, the \ac{gwd}~\cite{Givens1984} was used, as originally suggested by~\cite{yang2016metrics}.
The methods' parameters were chosen according to the parameters of the simulation. The number of iterations $l_{max}$ for \ac{vbrm} was set to $10$, as suggested in the original work~\cite{tuncer2021random}. 
The parameter controlling the rate of change of the shape was set to $0.98$.
For the \ac{rm} tracker, the scalar certainty was decayed by $\exp(-\frac{1}{1.6})$ for the moving object, and by $\exp(-\frac{1}{10})$ for the stationary one. 
These values were tuned to yield accurate tracking results and avoid divergence of the tracking results.
The number of particles for the \ac{memrbpf} was $50$ as suggested in the original article~\cite{steuernagel2025extended}.
As suggested in~\cite[Section IV-A]{gramsch2024batch}, for the \ac{mem}-based reference methods, we add a parameter $\psi$ to threshold the variance of the semi-axes to avoid the Gaussian distribution from reaching into negative values. The variances of the semi-axes are cut off at $(\psi\cdot\mat{l})^2$.
This value was selected as $0.4$ for the \ac{memekf}, as suggested in~\cite{gramsch2024batch}. For the \ac{memeif}, it was chosen even lower as $0.15$, leading to improved results in our experiments.
Our proposed algorithm does not require heuristic pruning of the axis variance, since the log-normal distribution only supports positive semi-axis lengths by nature.

\begin{figure}[t]
    \centering
    \newcommand\quantwidth{0.78}
    \subfloat[Moderate measurement data.]{\includegraphics[width=\quantwidth\columnwidth]{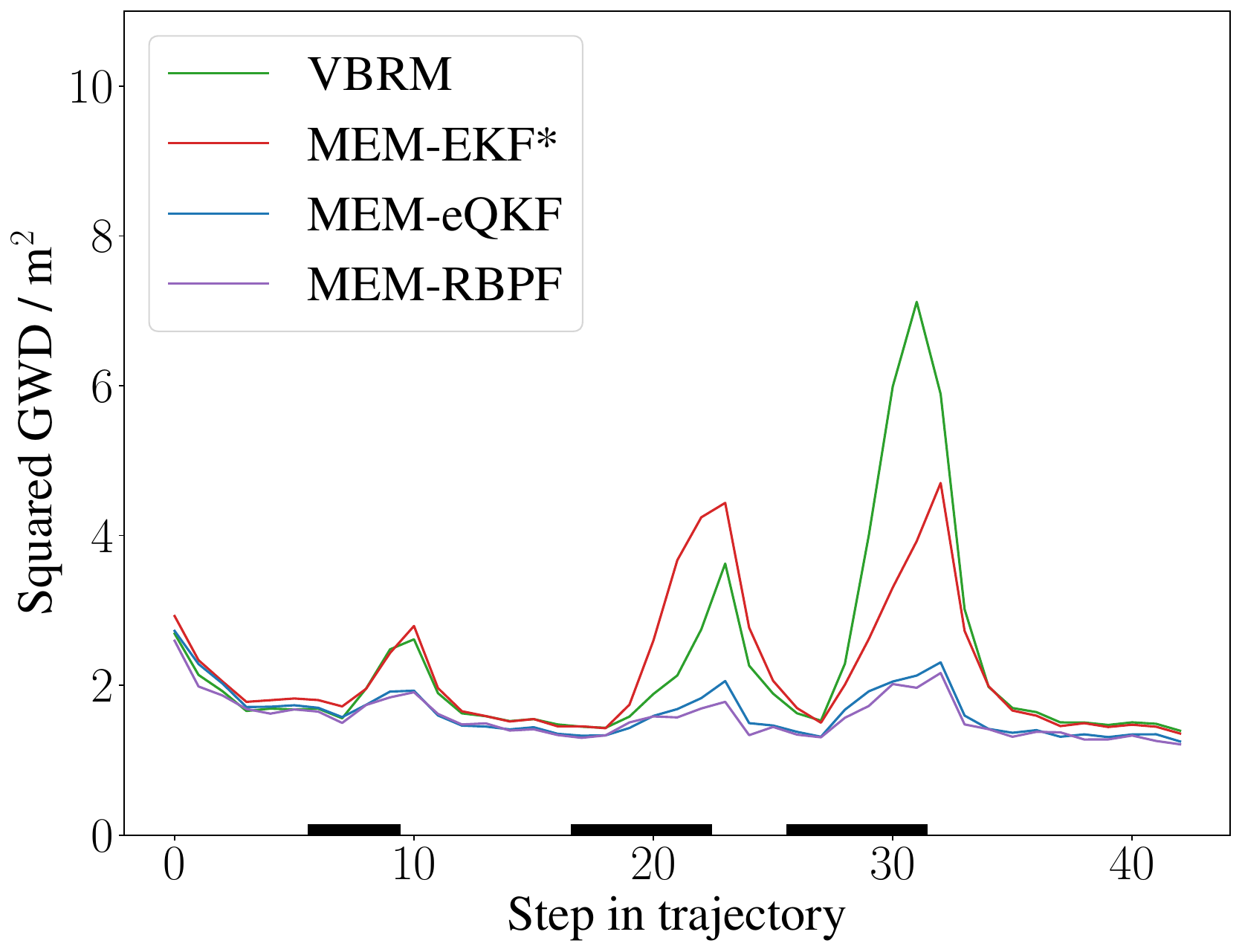}\label{fig:results:moderate}}\hfill
    
    \subfloat[Noisy measurement data.]{\includegraphics[width=\quantwidth\columnwidth]{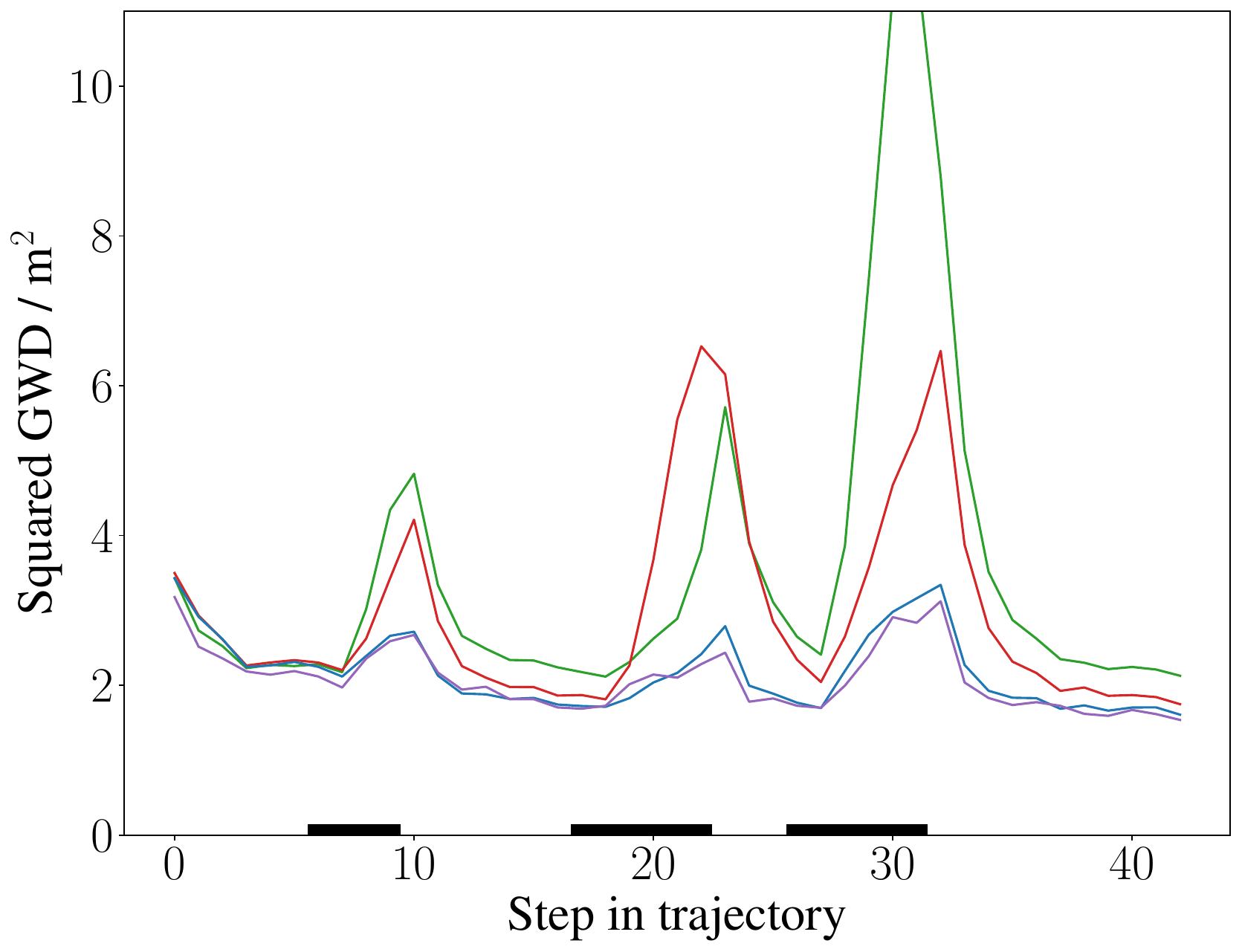}\label{fig:results:noisy}}\hfill
    
    \subfloat[Sparse measurement data.]{\includegraphics[width=\quantwidth\columnwidth]{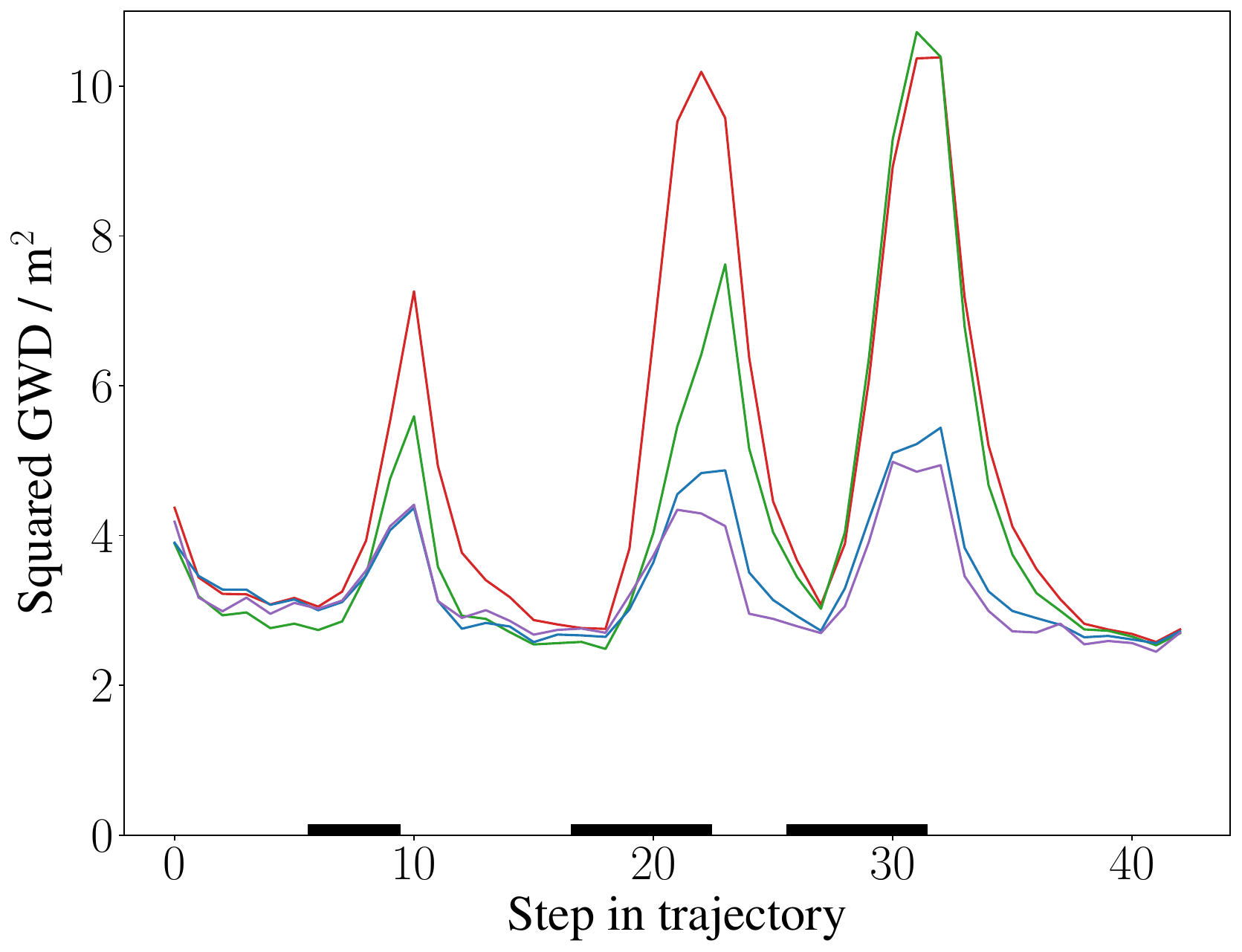}\label{fig:results:sparse}}\hfill
    \caption{
    Results of the evaluation for three different measurement settings, each averaged over $500$ Monte Carlo runs. 
    Black bars indicate time steps during which the object turned. 
    }
    \label{fig:results:quant}
\end{figure}

\begin{table}[t]
    \centering
    \caption{
    Mean error squared \ac{gwd} for all trackers on 500 Monte Carlo runs of each of the three scenarios. 
    Batch-based methods are separated in their own section.
    }
    \label{tab:results:means}
    \begin{tabular}{l|ccc}
    Method & Moderate & Noisy & Sparse  \\ \hline
MEM-EKF*~\cite{yang2019tracking}  & $\SI{2.185}{\square\metre}$ & $\SI{2.974}{\square\metre}$ & $\SI{4.738}{\square\metre}$ \\
VBRM~\cite{tuncer2021random}  & $\SI{2.242}{\square\metre}$ & $\SI{3.529}{\square\metre}$ & $\SI{4.143}{\square\metre}$ \\
MEM-RBPF~\cite{steuernagel2025extended}  & $\SI{1.576}{\square\metre}$ & $\SI{2.082}{\square\metre}$ & $\SI{3.277}{\square\metre}$ \\
MEM-eQKF (ours)  & $\SI{1.639}{\square\metre}$ & $\SI{2.176}{\square\metre}$ & $\SI{3.388}{\square\metre}$ \\ \hline
RM~\cite{feldmann2010tracking}  & $\SI{3.494}{\square\metre}$ & $\SI{4.377}{\square\metre}$ & $\SI{10.194}{\square\metre}$ \\
MEM-EIF~\cite{gramsch2024batch}  & $\SI{2.387}{\square\metre}$ & $\SI{3.014}{\square\metre}$ & $\SI{4.802}{\square\metre}$ \\
Batch MEM-eQKF (ours)  & $\SI{1.899}{\square\metre}$ & $\SI{2.489}{\square\metre}$ & $\SI{3.789}{\square\metre}$ \\
    \end{tabular}
    
\end{table}

For all experiments, the expected value of the prior of the state was passed to the tracker along with the corresponding covariance.
The ground truth was sampled from this distribution, without the result being known to the tracker.
An implementation of the method as well as the presented experiments including the real-world radar data is available online\footnote{\url{https://github.com/Fusion-Goettingen/}}.

\subsection{Moving Target}\label{sec:simulation:general}
In the first set of experiments, the object followed the path shown in Fig.~\ref{fig:traj_overview}, modeled to follow the evaluation trajectory used in~\cite{yang2019tracking}. 
The shape prior had expected value $\SI{5}{\metre}$ and $\SI{2}{\metre}$ for the semi-axis, with covariance $\mat{I}_{2\times2}$. 
The kinematic prior covariance matrix was $\diag(2, 2, 0.5, 0.5)$. 
The process noise for the kinematics was $1$ for the position and $2$ for the velocity. The target orientation was aligned with the velocity, which was encoded with variance $0.1$ radians in all trackers.
The object performs three turns, violating the employed constant velocity motion model. Employing a more complex, nonlinear model could improve tracking performance, however, it is of great interest to evaluate how the trackers perform when the motion model is temporarily violated. This is because a key challenge of \ac{eot} is dealing with the fact that measurement locations can shift due to either kinematic or shape changes of the object. 

A total of 500 Monte Carlo runs were performed for three settings. These settings differ in two parameters, the expected number of measurements $\lambda$ and the measurement noise covariance matrix $\mat{R}$.
The realizations of these values were as follows:
\begin{itemize}
    \item Moderate: {$\lambda=12$, $\mat{R}={\rot{\frac{\pi}{4}}\diag\left(\frac{3}{2}, \frac{2}{3}\right)\rot{\frac{\pi}{4}}^\tp}$}
    \item Noisy: $\lambda=12$, $\mat{R}={\rot{\frac{\pi}{4}}\diag\left(3, 1\right)\rot{\frac{\pi}{4}}^\tp}$
    \item Sparse: $\lambda=6$, $\mat{R}={\rot{\frac{\pi}{4}}\diag\left(\frac{3}{2}, \frac{2}{3}\right)\rot{\frac{\pi}{4}}^\tp}$
\end{itemize}
Anisotropic covariance matrices are used to simulate sensors of interest for \ac{eot}, e.g., radar. The number of measurements in each step was drawn from a Poisson distribution with mean $\lambda$.

The mean tracking error over time of the methods measured in \ac{gwd} is shown in Fig.~\ref{fig:results:quant}. Additionally, the overall averages across all runs of each of the settings are summarized in Table~\ref{tab:results:means}.
As mentioned above, situations in which the motion of the target is not accurately predicted cause a significant increase in error. This effect is aggravated in the sparse and noisy settings.
Here, the effective difference between methods is best visible.
The sampling-based \ac{memrbpf} consistently performs best throughout all settings. The closed-form \ac{rm} algorithm is affected most, whereas the optimization-based \ac{vbrm} can recover noticeably faster. 
Our proposed method exhibits robust behavior, almost matching the performance of the sampling-based \ac{rbpf}. This leads to a significant reduction in error compared to the \ac{memekf}.
This can also be seen in the qualitative results presented in Fig.~\ref{fig:traj_overview}.
The summary presented in Table~\ref{tab:results:means} confirms the observations visible in Fig.~\ref{fig:results:quant}: In all settings, the mean error of the \ac{memrbpf} is lowest, closely followed by the proposed method. A noticeable gap between the error of these two methods and all other reference methods, including the optimization-based \ac{vbrm}, can be observed.

\begin{table}[t]
    \centering
    \caption{
    Mean orientation estimation error throughout 500 Monte Carlo runs of the moderate measurement setting. 
    Batch-based methods are separated in their own section.
    }
    \label{tab:results:orientation}
    \begin{tabular}{l|c}
    Method & Orientation Error (rad) \\ \hline
    MEM-EKF*~\cite{yang2019tracking}  & 0.267 \\
    VBRM~\cite{tuncer2021random}  & 0.248 \\
    MEM-RBPF~\cite{steuernagel2025extended}  & 0.191 \\
    MEM-eQKF (ours)  & 0.205 \\ \hline
    RM~\cite{feldmann2010tracking}  & 0.299 \\
    MEM-EIF~\cite{gramsch2024batch}  & 0.219 \\
    Batch MEM-eQKF (ours)  & 0.224 \\
    \end{tabular}
\end{table}

\begin{figure}
    \centering
    \includegraphics[width=0.78\columnwidth]{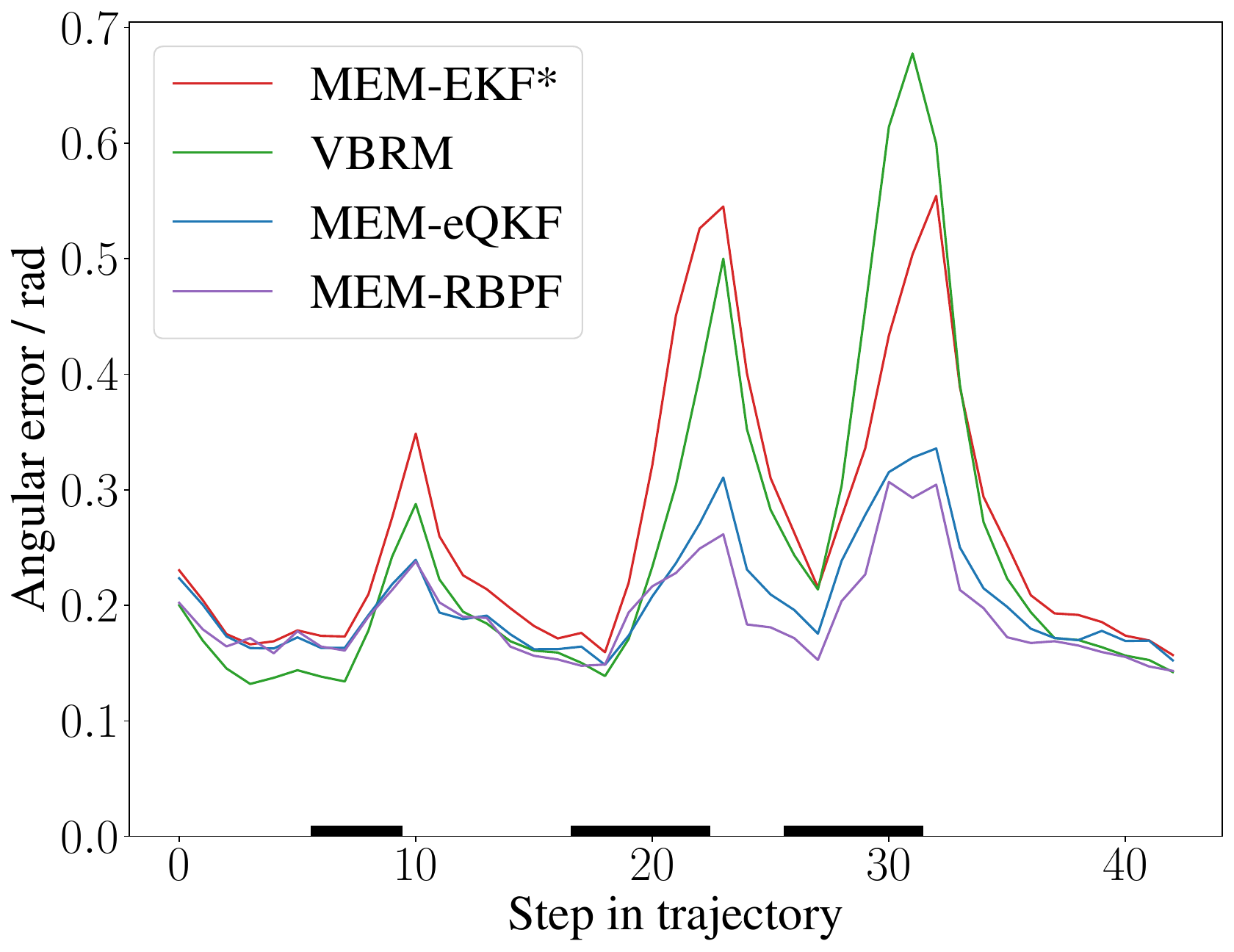}
    \caption{
    Visualization of the orientation error of the iterative methods, average over 500 Monte Carlo runs. Measurement parameters were chosen according to the ``moderate'' setting.
    }
    \label{fig:results:theta}
\end{figure}

Beyond the comparison in \ac{gwd}, we compare the orientation estimation quality in Table~\ref{tab:results:orientation} and in Fig.~\ref{fig:results:theta}. 
Generally, the orientation error and the \ac{gwd} clearly correlate as expected. The \ac{rbpf}, which has particular emphasis on the orientation estimation in its design, performs best. 
This is also immediately visible in Fig.~\ref{fig:results:theta}, where the \ac{memrbpf} consistently yields an error curve below the other methods, with the exception of a short period in the beginning where the \ac{vbrm}~\cite{tuncer2021random} performs better.
Noticeably, the proposed \ac{memeqkf} performs decidedly better than the \ac{memekf}, despite the mathematical similarity in their respective orientation estimation procedures. 
As can be seen from Fig.~\ref{fig:results:theta}, this is particularly emphasized during the challenging turns.
We attribute the improved performance compared to the reference tracker to two factors: 
First, the improved semi-axis estimation procedures allow for better orientation estimation. 
Second, the fact that by approximating the correlations between orientation and axes as zero, fewer errors are introduced.

\begin{table}[t]
    \centering
    \caption{
        Mean runtime of the compared methods across the $500$ Monte Carlo runs with $20$ measurements per step. 
        Batch-based methods are separated in their own section.
    }
    \label{tab:runtime}
    \begin{tabular}{l|c}
        Method & Mean Runtime \\ \hline
MEM-EKF*~\cite{yang2019tracking}  & $\SI{7.772}{\milli\second}$ \\
VBRM~\cite{tuncer2021random}  & $\SI{13.422}{\milli\second}$ \\
MEM-RBPF~\cite{steuernagel2025extended}  & $\SI{8.181}{\milli\second}$ \\
MEM-eQKF (ours)  & $\SI{7.896}{\milli\second}$ \\ \hline
RM~\cite{feldmann2010tracking}  & $\SI{1.467}{\milli\second}$ \\
MEM-EIF~\cite{gramsch2024batch}  & $\SI{1.919}{\milli\second}$ \\
Batch MEM-eQKF (ours)  & $\SI{2.057}{\milli\second}$ \\
    \end{tabular}
\end{table}
\begin{figure}[t]
    \centering
    \includegraphics[width=0.78\columnwidth]{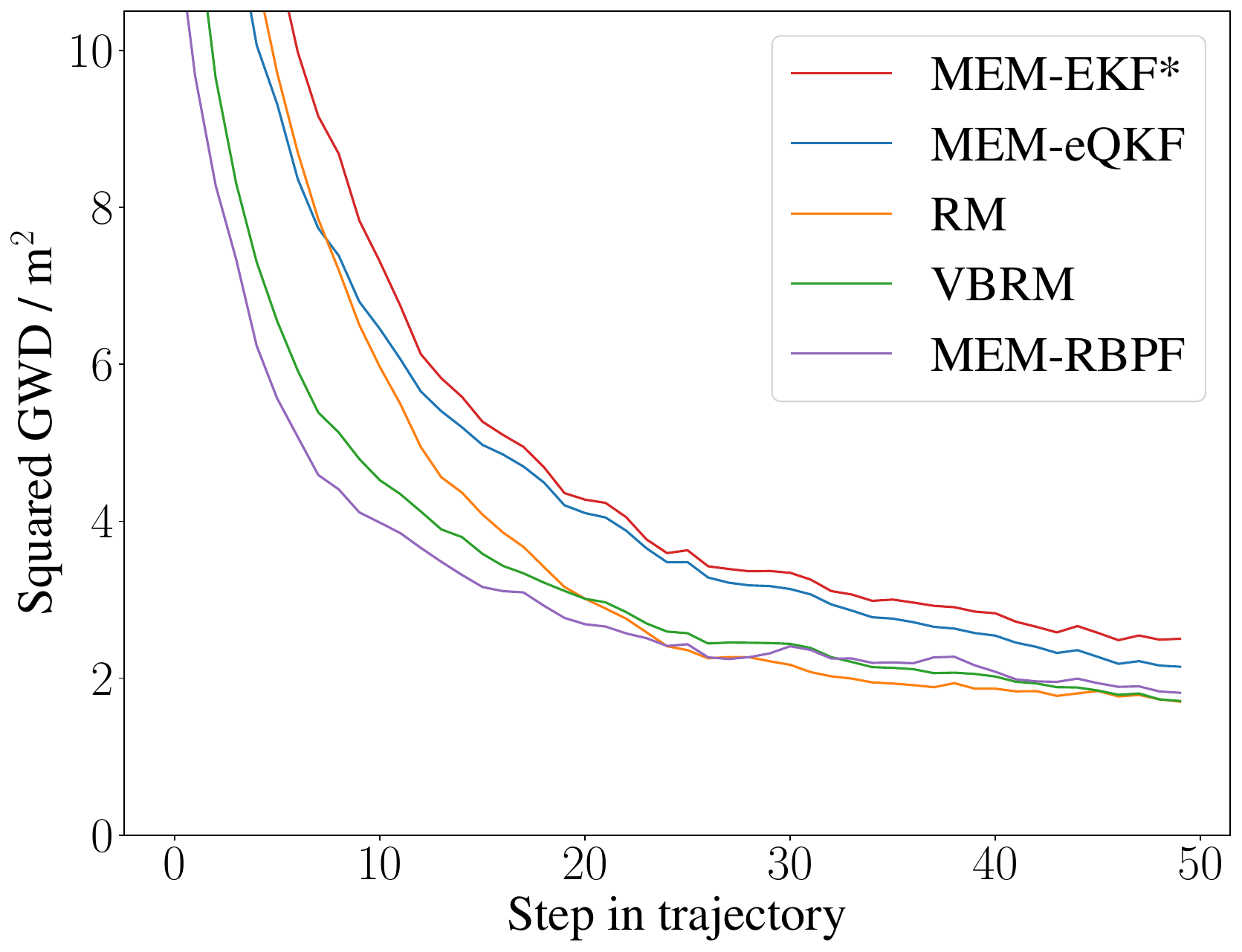}
    \caption{
    Estimation error in \acrshort{gwd} for a stationary target with a single measurement per step, averaged over $500$ Monte Carlo runs.
    }
    \label{fig:results:single}
\end{figure}
The runtime of the methods is compared in Table~\ref{tab:runtime}, and also below in Fig.~\ref{fig:results:runtime_over_meas}. All methods were implemented in Python and run on the same system using an Intel i5-1145G7 processor.
The results of this confirm that the proposed closed-form method is significantly faster than the optimization-based \ac{vbrm} and the sampling-based \ac{memrbpf}. 
Differences between the MEM-EKF* and the proposed method are minimal due to their similarity.
Overall, the batch-based \ac{rm} approach is nevertheless even more computationally efficient, outperforming all other methods.

\subsection{Single Measurement per Step}\label{sec:simulation:singlemeas}
\begin{figure}[t]
    \centering
    \includegraphics[width=0.78\columnwidth]{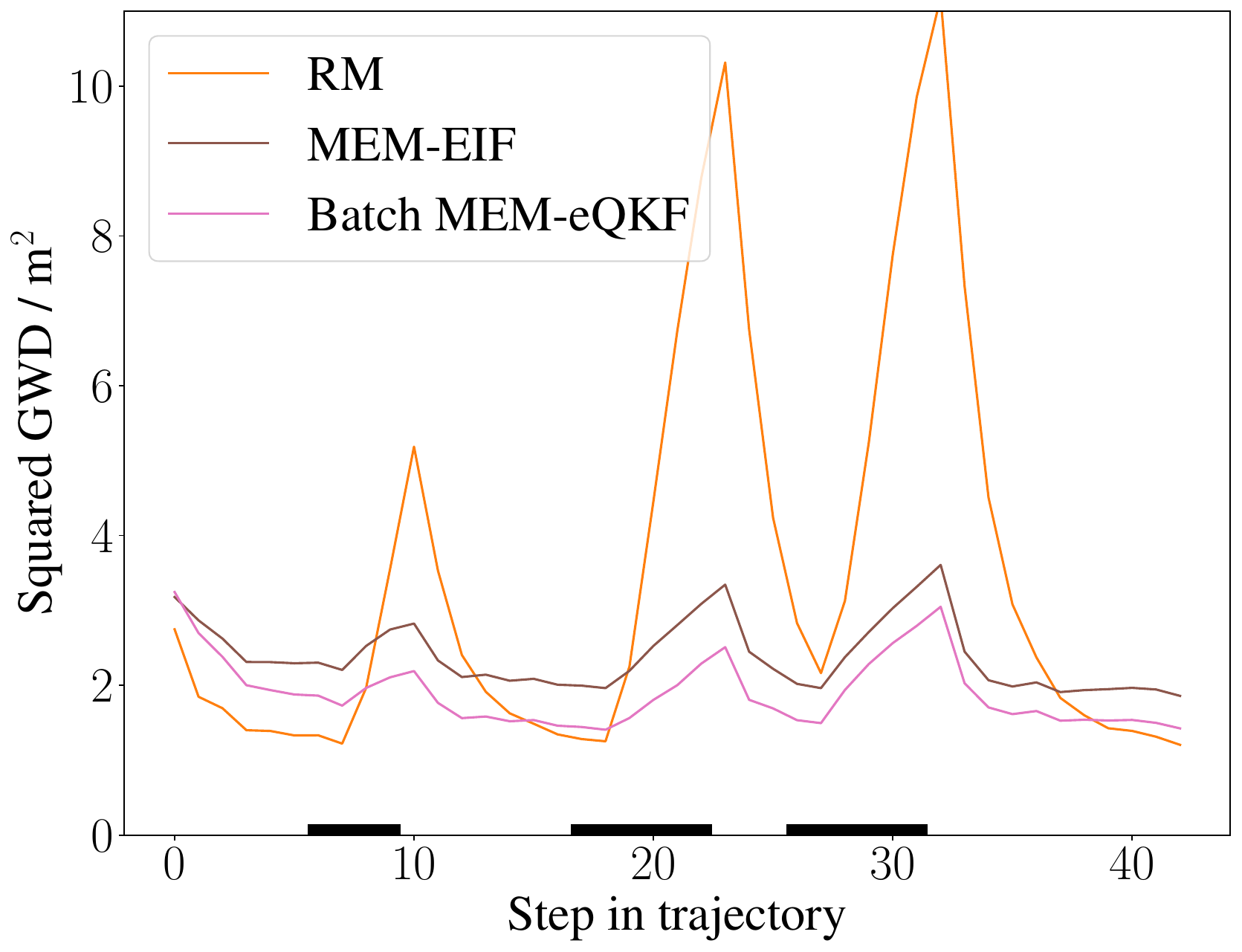}
    \caption{
    Results of the three batch-based algorithms in the moderate measurement setting, averaged over 500 Monte Carlo runs. Black bars indicate time steps during which the object turned. 
    }
    \label{fig:results:batch:moderate}
\end{figure}
As a second simulation study, we evaluated the results of the methods for the setting of a stationary target with expected size $8\times4$ meters emitting exactly one measurement per time step. The measurement noise was $\mat{R}=\mat{I}_{2\times2}$. Results are visualized in Fig.~\ref{fig:results:single}.
All trackers were initialized with variance of $0.1$ for the position, but the initial orientation uncertainty was $\pi$ and the initial covariance of the axis lengths $\diag(4, 2)$. 
This means the kinematics were more or less entirely known to the trackers, but due to the strong noise on the shape prior, the convergence behavior of the extent estimation of each method can be studied.
As for a single measurement, the batch approximations for the proposed algorithm and the \ac{memekf} are not needed, these are not included in the plot.
Furthermore, it is important to note that for the \ac{rm} approach to converge at all, the decay of the scalar certainty parameter of the inverse Wishart distribution modeling the shape needs to be vastly reduced compared to the previous experiment.

With this in mind, the \ac{rm} approach converges to the lowest error of all methods, albeit after a slower convergence phase than the \ac{vbrm} and \ac{memrbpf}. These two methods converge to a similarly low error, with the \ac{memrbpf} converging the fastest, likely due to the fact that it simultaneously evaluates multiple hypotheses about the orientation.
The \ac{memekf} and \ac{memeqkf} converge to slightly higher overall error.
As in the previous experiment, the \ac{memeqkf} outperforms the \ac{memekf}, both in final error and throughout the convergence phase.
This experiment demonstrated that~\eqref{eq:meascenter:stream} is a suitable approach for dealing with a stream of measurements.

\subsection{Batch-based Variant}\label{sec:simulation:batch}
The batch-based variant proposed in Section~\ref{sec:batch} has been analyzed using similar experiments as described above.
Tables~\ref{tab:results:means},~\ref{tab:results:orientation}, and~\ref{tab:runtime} include separate sections containing aggregated mean results for the three batch-based algorithms.
Furthermore, the mean error over 500 Monte Carlo runs of the moderate measurement setting for the batch-based methods is presented in Fig.~\ref{fig:results:batch:moderate}.

Initially, the \ac{rm} algorithm performs best, however, it is severely affected by the turns, leading to very large spikes in error.
These are significantly less pronounced for the two \ac{mem}-based algorithms. Out of these two, our proposed method consistently produces more accurate results compared to the \ac{memeif}.

\begin{figure}[t]
    \centering
    \includegraphics[width=0.95\columnwidth]{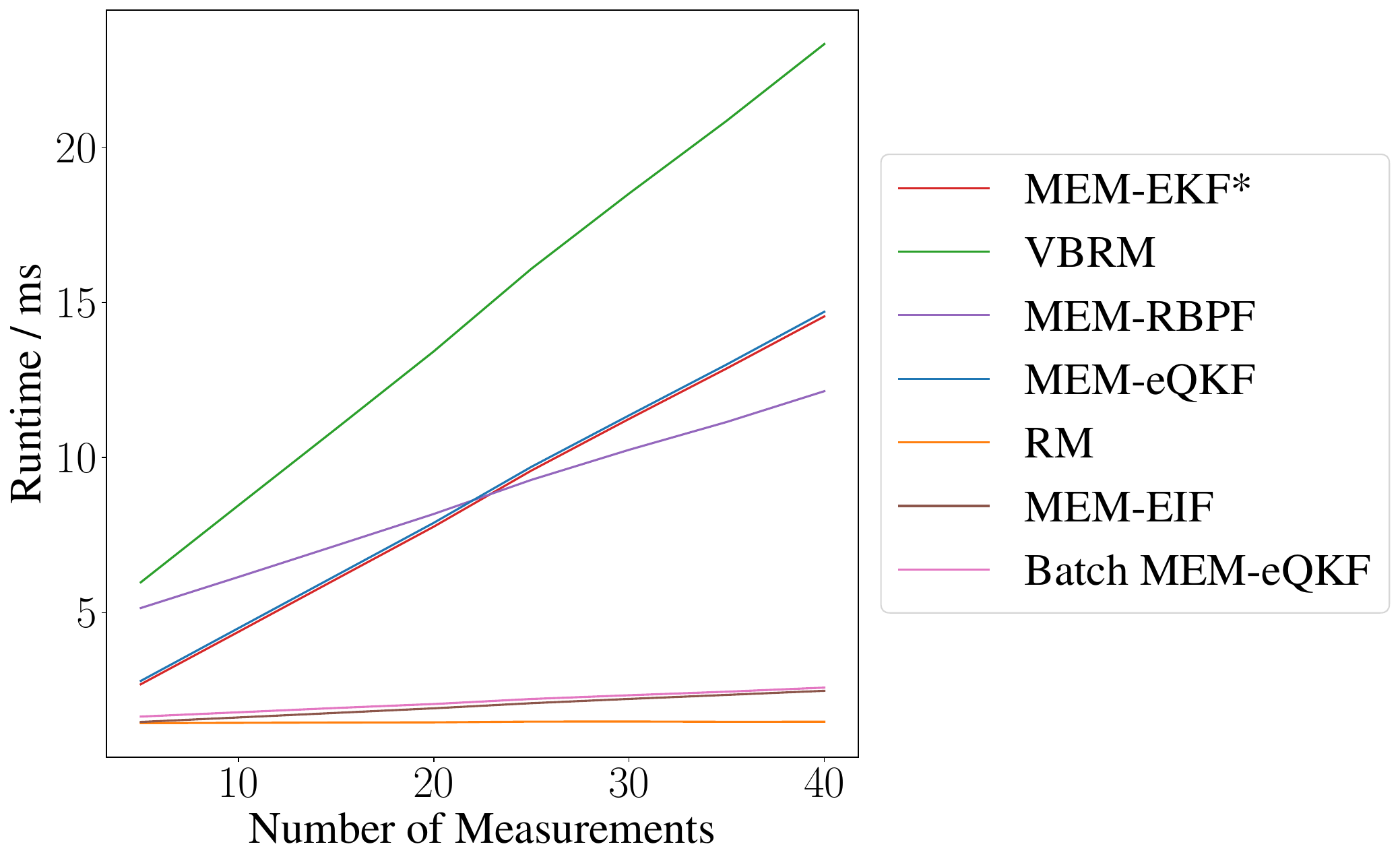}
    \caption{
    Runtime of the evaluated methods for different number of processed measurements, averaged over $50$ Monte Carlo runs of different settings with increasing measurement count.
    }
    \label{fig:results:runtime_over_meas}
\end{figure}

Finally, Fig.~\ref{fig:results:runtime_over_meas} presents an analysis of the runtime of the evaluated methods over the number of measurements. For this experiment, the same settings as in the moderate measurement settings have been used. However, the number of measurements in each time step was varied between runs, and in each step fixed to the respective number, rather than drawn from a Poisson distribution as before. 
The efficiency improvement of the proposed batch-based variant can be clearly seen in this figure. While it does not quite match the computational efficiency of the \ac{rm} approach, it significantly outperforms all algorithms processing measurements sequentially. The similar \ac{memeif}~\cite{gramsch2024batch} yields almost identical runtime.

In summary, the batch-based variant of the proposed \ac{memeqkf} yields a tracker of high computational efficiency. Still, the tracking results are of high quality, leading to the fact that the tracker nevertheless outperforms state-of-the-art sequential algorithms, as seen in Table~\ref{tab:results:means}.

\subsection{Ablation Study}\label{sec:simulation:ablation}
In addition to the evaluation comparing the proposed approach with several state-of-the-art algorithms, additional experiments have been carried out to analyze behavior of the method in special cases.
Four aspects have been analyzed: 
First, the pseudo-measurement independence assumption applied in the derivation of the batch-based axis update.
Second, the behavior in the corner case that the object is circular, meaning that the orientation cannot be estimated.
Third, the effect of strong turn rates on the filter and fourth, the susceptibility of the method to incorrect covariance estimates.

\subsubsection{Pseudo-Measurement Independence}~\label{sec:simulation:pseudomeas-independence}

The batch-based update of the semi-axis lengths makes use of the batch innovation covariance~\eqref{eq:batch:axis:innov-cov}. This implicitly approximates all pseudo-measurements as independent, despite correlations existing due to the dependency of the state.
An equivalent formulation to~\eqref{eq:batch:axis:innov-cov} would be to formulate a stacked update with a block-diagonal covariance matrix for the stacked measurements.

\begin{figure}
    \centering
    \includegraphics[width=\columnwidth]{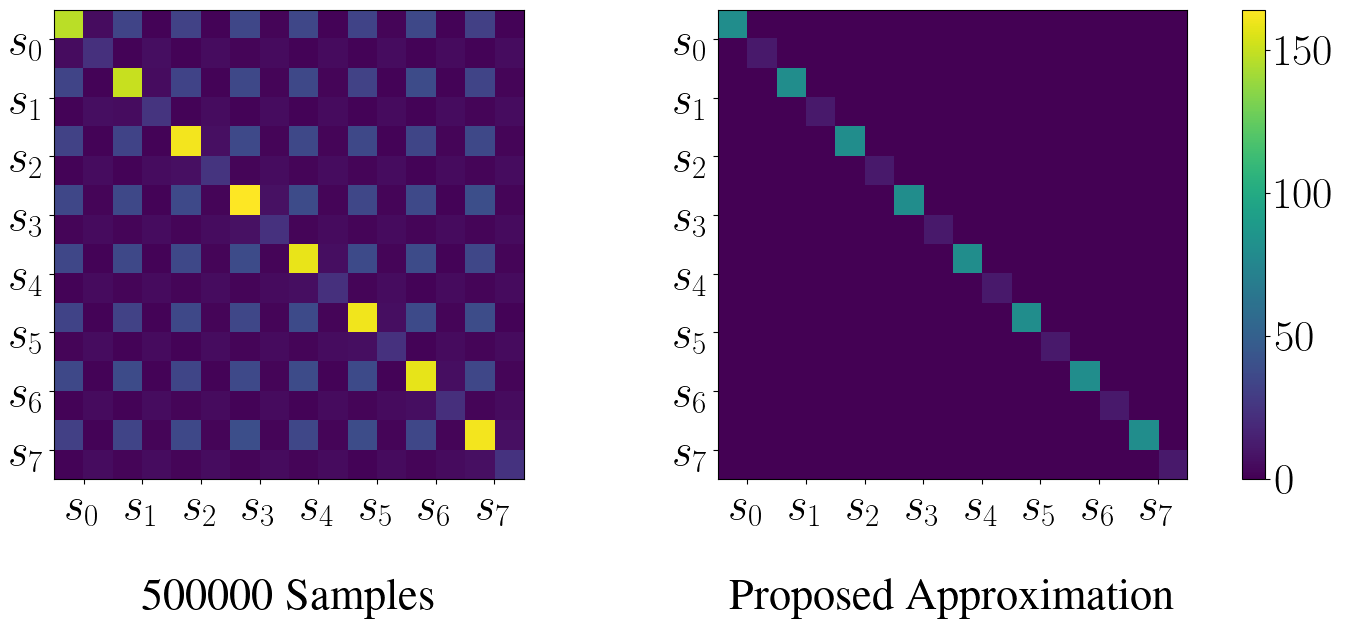}
    \caption{
        Comparison of the numerically determined covariance of the batch of pseudo-measurements (left) and the employed approximation (right). 
    }
    \label{fig:ablation:pseudo_meas_cov}
\end{figure}

In Fig.~\ref{fig:ablation:pseudo_meas_cov}, two covariance matrices are visualized: On the left, the actual (numerical) covariance computed from $500000$ samples, and on the right the approximation used in the batch-based \ac{memeqkf}, visualized in the explicit block-diagonal form to show the lack of between-measurement correlations. 
For the sampled covariance, object and noise parameters were fixed, and samples of batches of eight measurements were drawn accordingly. The $16\times16$ covariance matrix was then computed from the pseudo-measurements computed as described above. 

The result clearly shows that correlations between the individual pseudo-measurements exist, which are not modeled in the approximation used in the filter, shown on the right. 
The linearized approximation is not perfectly accurate for the main diagonal.

\begin{figure}
    \centering
    \includegraphics[width=0.9\columnwidth]{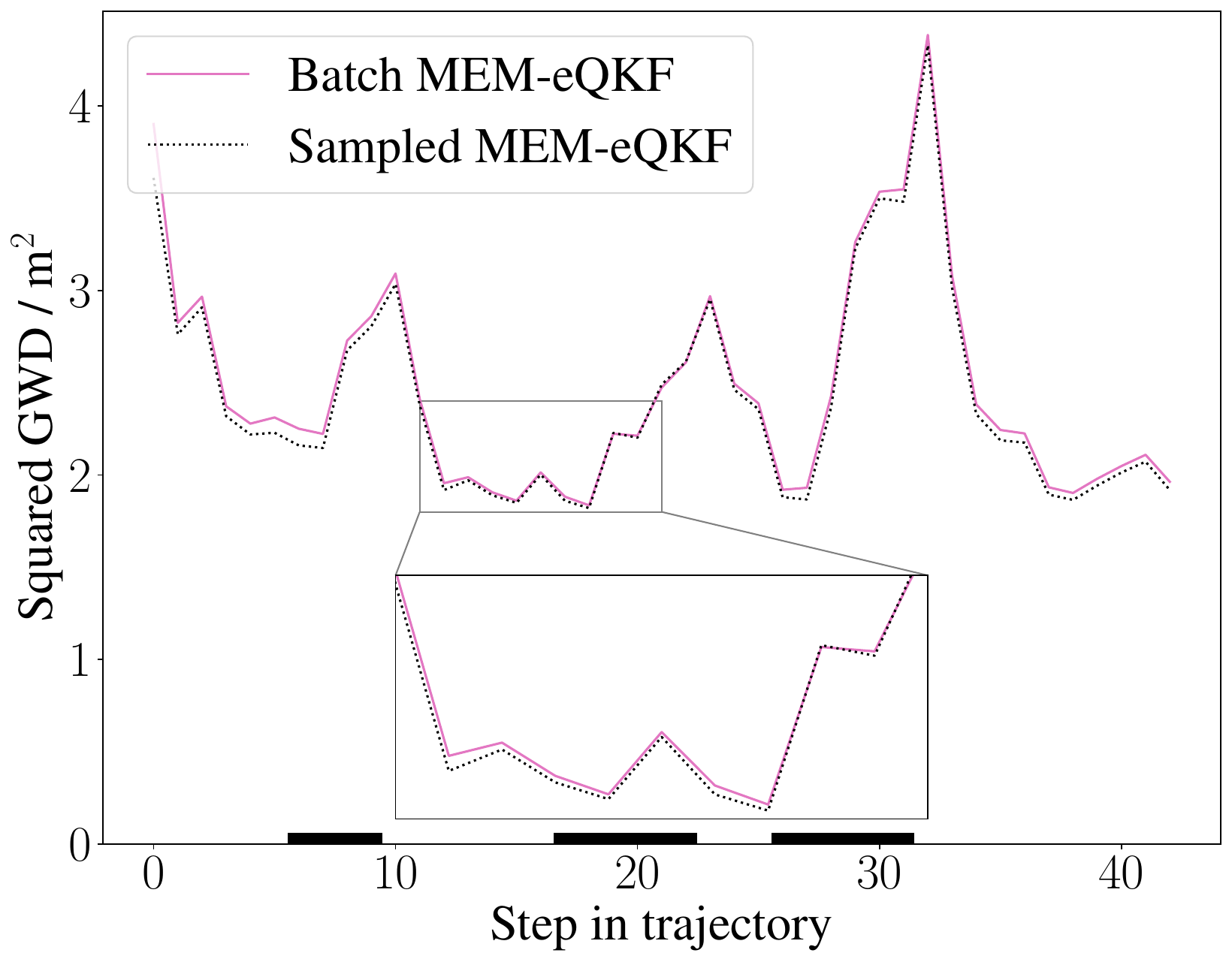}
        \caption{
        Average error across 100 Monte Carlo runs comparing the approximation based Batch \acrshort{memeqkf} with a variant employing a numerical sampling-based approach for the computation of the batch pseudo-measurement covariance. The setting used was the ``noisy'' measurement parameterization.
    }
    \label{fig:ablation:sampled_cov_tracking}
\end{figure}

The key question is, however, the impact of the independence assumption on the tracking performance. To evaluate this, the Batch \ac{memeqkf} was compared to a second version in which the above covariance was sampled as described above. 
For this sampling, the prior of the tracker was used in each measurement update to compute $10000$ new samples from which the numerical sample covariance was determined. Based on this, the explicit stacked measurement update was computed. 
Average results over 100 Monte Carlo runs are shown in Fig.~\ref{fig:ablation:sampled_cov_tracking}. 
The ``noisy'' measurement settings described above have been used for the evaluation.
Barely any difference between the two algorithms is visible. Closer inspection reveals the sampling version to yield an average error of $\SI{2.42}{\square\metre}$, compared to the error of $\SI{2.46}{\square\metre}$ achieved by the proposed batch-based version that makes use of the approximation regarding independence. Overall, the error would be reduced by less than $2\%$ when eliminating this approximation. Visual inspection of the qualitative results confirms that the two perform nearly identically.

In summary, experiments have shown that while the approximation regarding independence made in~\eqref{eq:batch:axis:innov-cov} indeed ignores existing correlations between pseudo-measurements, this does not affect the error in any significant manner.

\begin{figure}[t]
    \centering
    \subfloat[Qualitative example.]{\includegraphics[width=1\columnwidth]{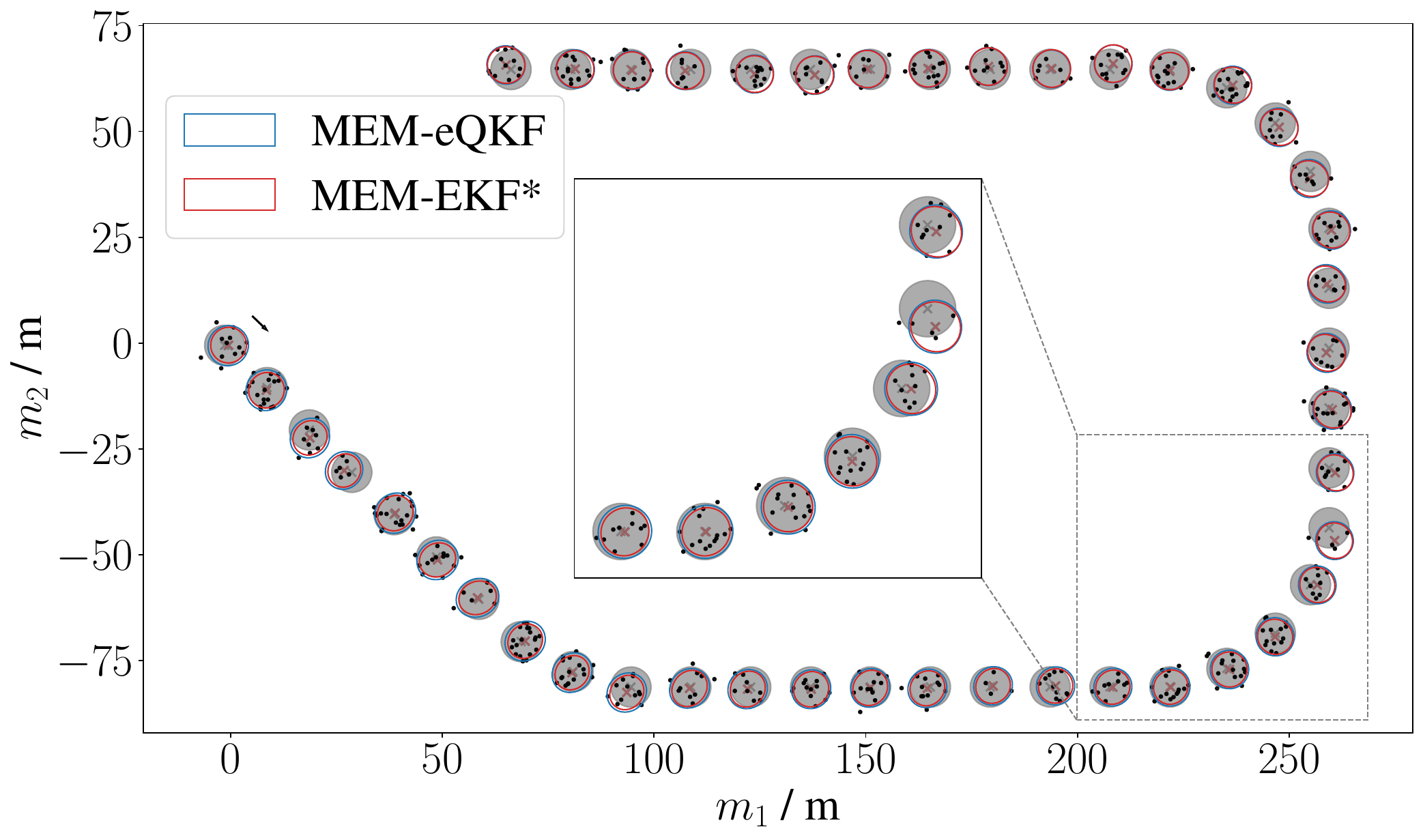}\label{fig:ablation:circular:example}}\hfill
    
    \subfloat[Quantitative results.]{\includegraphics[width=0.78\columnwidth]{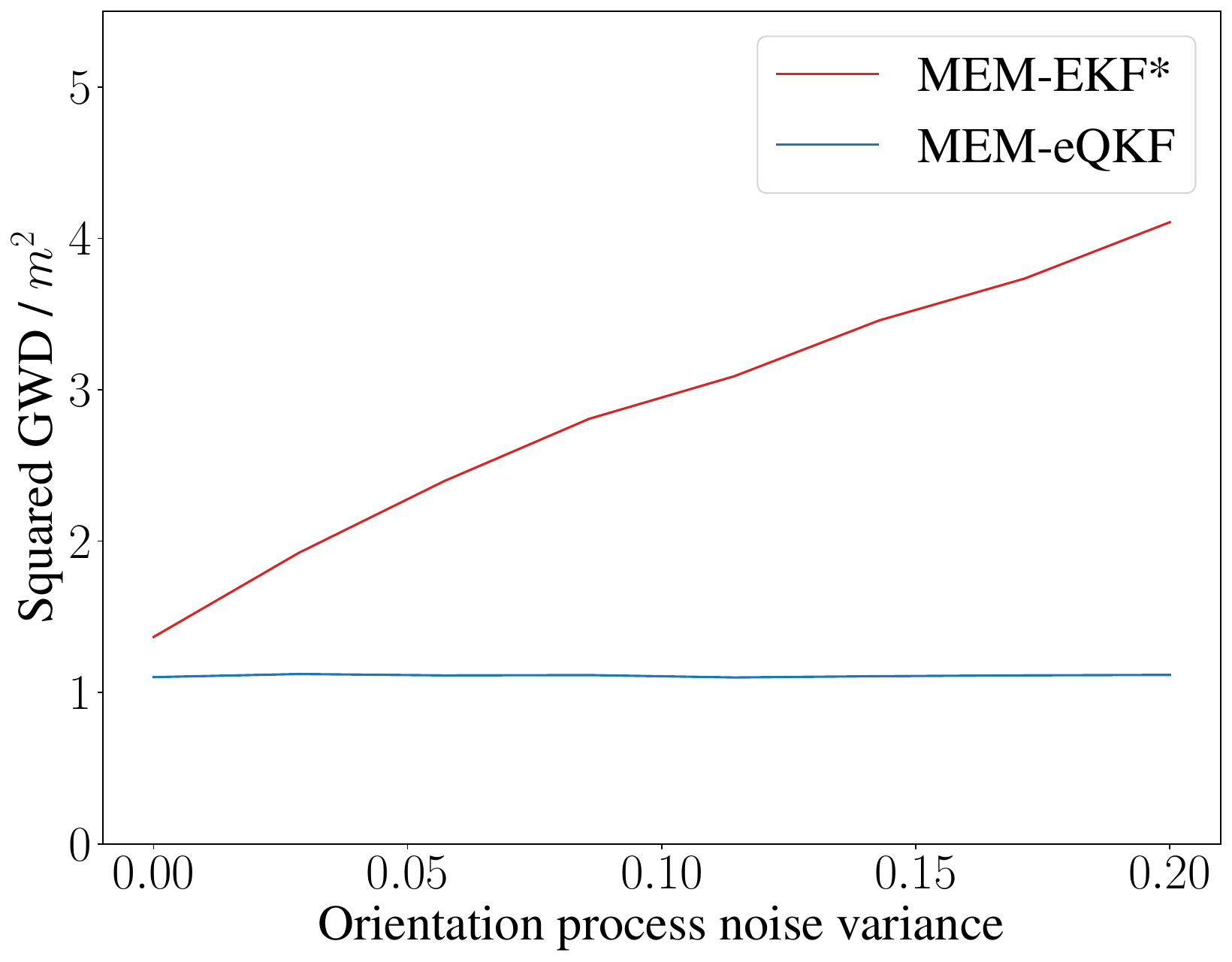}\label{fig:ablation:circular:quant}}\hfill
    \caption{
        Comparison of the \ac{memekf}~\cite{yang2019tracking} and the proposed \ac{memeqkf} for tracking a circular object. 
        The measurement parameters were chosen according to the ``moderate'' setting. 
        For the quantitative results, the average error for different settings of the orientation process noise variance is shown, averaged over 500 Monte Carlo runs for each choice of noise.
    }
    \label{fig:ablation:circular}
\end{figure}

\subsubsection{Circular Objects}

If the object is (close to) circular, i.e., isotropic, the orientation is not defined anymore, since rotating a circle does not affect its geometric form at all.
Therefore, it can be reasonable to assume that in this case, the proposed decoupled strategy that explicitly makes use of the (unobservable) orientation collapses.
However, it is important to recognize that in this case, any orientation estimation error has no impact: This is again due to the very same fact that rotating a circle does not affect its geometry. 

Still, in order to confirm that the proposed algorithm does not fail under such circumstances, experiments were conducted. These were designed exactly as the experiments in the previous sections, with the ``moderate'' noise settings. However, here, the ground truth expected semi-axis size was set to $\SI{4}{\metre}$ for both semi-axes, and the true object size was assured to be isotropic. 

The results, shown in Fig.~\ref{fig:ablation:circular}, confirm that the method indeed still works without requiring any further adaptation for the case that the object is isotropic. 
The quantitative results presented in Fig.~\ref{fig:ablation:circular:quant} show the \ac{memekf}~\cite{yang2019tracking} to not be as robust regarding the selection of the orientation process noise as the proposed \ac{memeqkf}. For a circular object, changes to the orientation make no difference to the resulting geometry. Nevertheless, as no knowledge about the fact that the object is circular is assumed a-priori, the process noise model used by the tracker must account for the fact that the target may be a rotating elliptical object. For the \ac{memekf} to yield accurate results, however, the lack of orientation process noise must be explicitly known by the tracker. Otherwise, the semi-axis lengths are severely underestimated.
Qualitatively, as seen in Fig.~\ref{fig:ablation:circular:example}, with low enough orientation process noise, both evaluated algorithms accurately estimate the object shape.

In summary, the evaluated corner case is of no concern for the proposed algorithm. In fact, related state-of-the-art methods can be outperformed in this case.

\begin{figure}[t]
    \centering
    \subfloat[Trajectory visualization.]{\includegraphics[width=0.8\columnwidth]{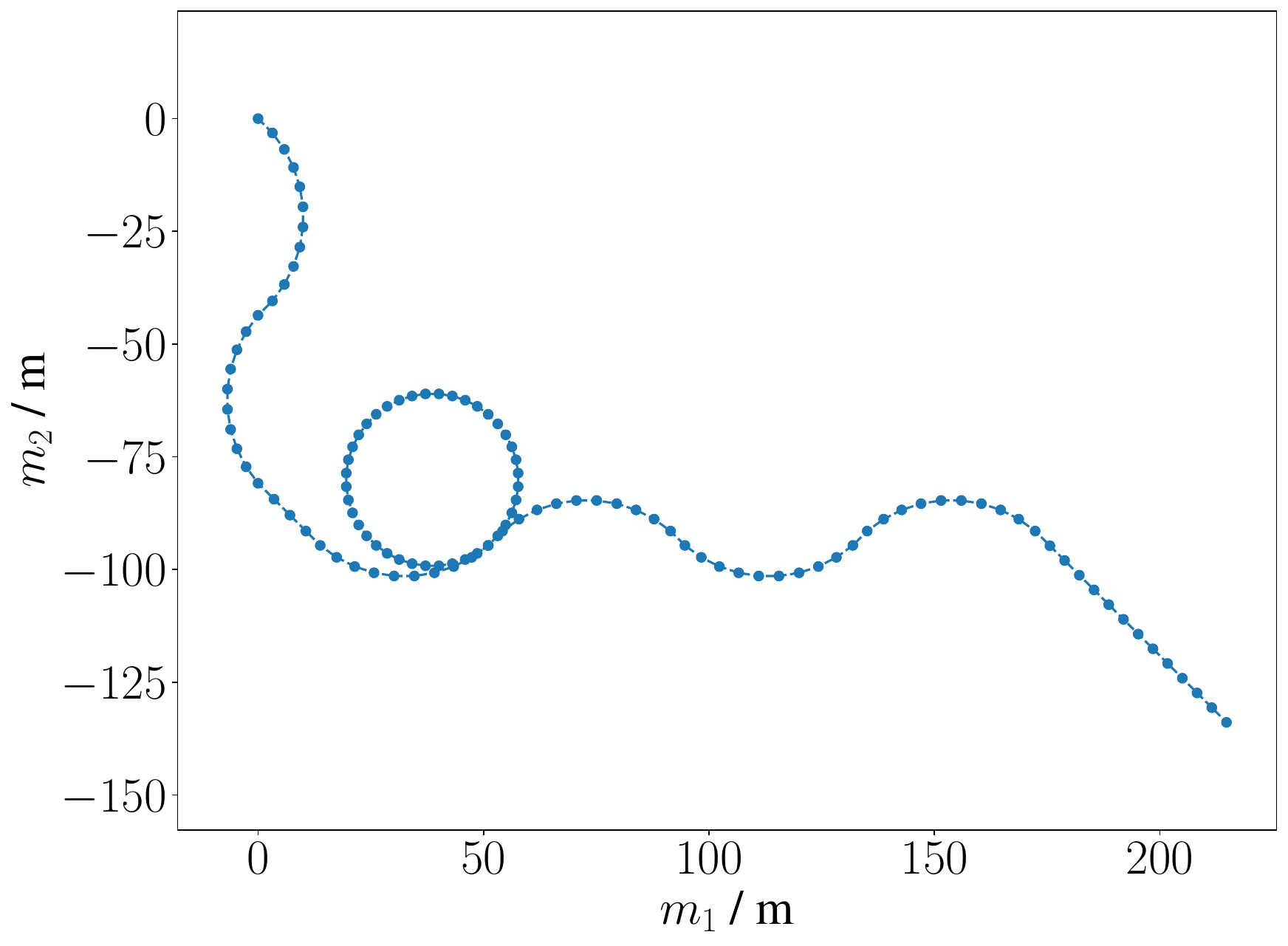}\label{fig:ablation:turnrate:example}}\hfill
    
    \subfloat[Quantitative results.]{\includegraphics[width=0.75\columnwidth]{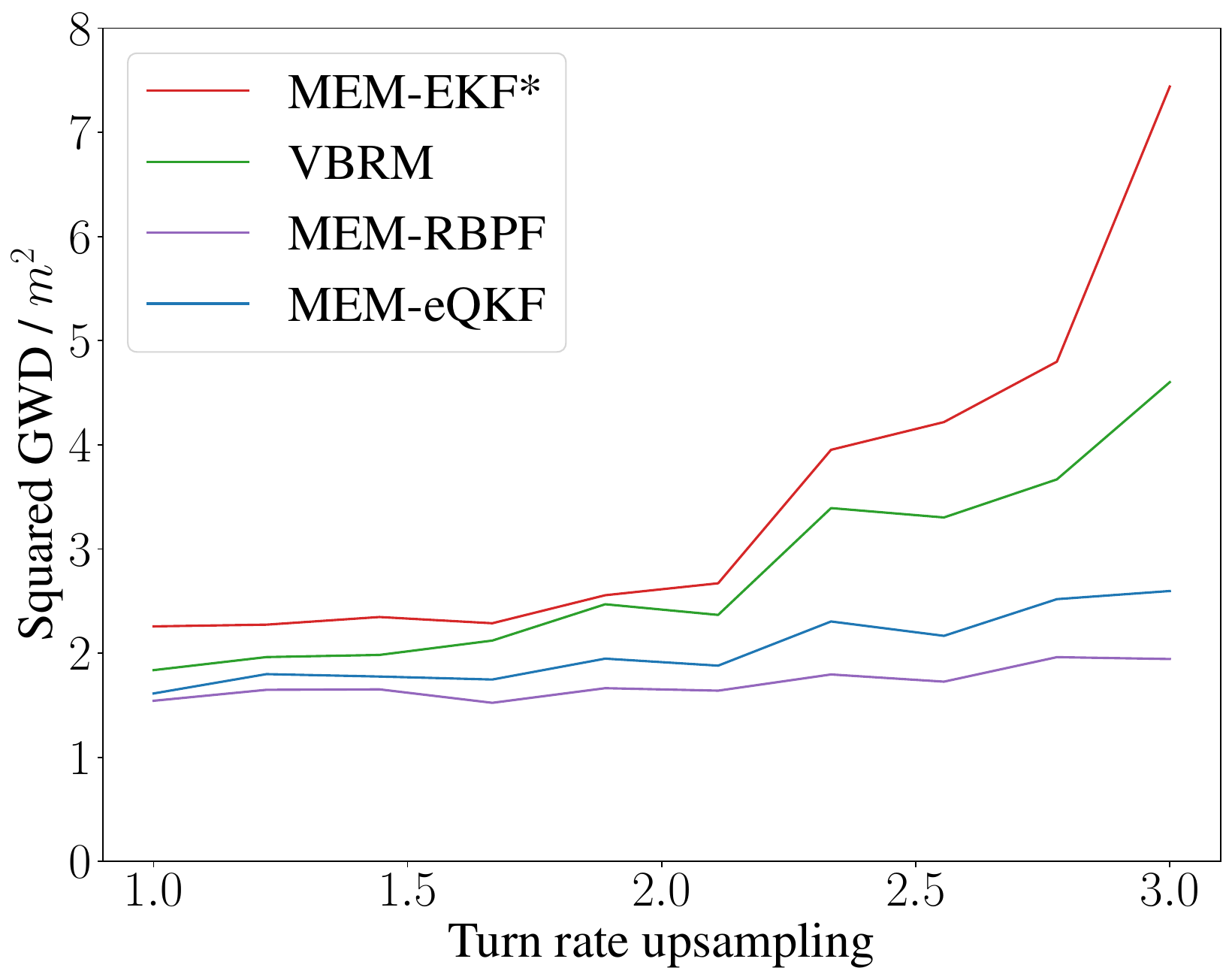}\label{fig:ablation:turnrate:quant}}\hfill
    \caption{
        Testing trajectory and results for the case that strong orientation changes. 
        The quantitative results are computed for different granularity of the true trajectory, implying different strength of turn rate of the object.
    }
    \label{fig:ablation:turnrate}
\end{figure}

\subsubsection{Strong Turn Rates}
To ensure that strong turn rates of the tracked object can be handled by the proposed algorithm, an ablation experiment has been designed to test such behavior.
A trajectory with many turns, visualized in Fig.~\ref{fig:ablation:turnrate:example}, was used for this. The original trajectory was sampled at a high rate, yielding overall low turn rates between time steps due to discretization. By increasing the upsampling, stronger turns between individual steps can be introduced, which needs to be accounted for by the trackers.

Results of this experiment using the four sequential \ac{eot} algorithms are presented in Fig.~\ref{fig:ablation:turnrate:quant}. The upsampling factor was linear increased from $1$ (no upsampling) to $3$ (effectively tripling the turn rate to be estimated). Moderate measurement settings as described above were used.
Results show that the \ac{vbrm} but also in particular the \ac{memekf} are affected by this. The proposed \ac{memeqkf} is affected, albeit to a significantly lower degree compared to any of the other deterministic filter.
The only filter that is barely affected is the \ac{memrbpf}, which by design handles stronger turns well due to its particle-based nature of the orientation representation.
In summary, the proposed algorithm handles strong turn rates significantly better than related algorithms.

\subsubsection{Incorrect Covariance Estimates}
For real-world applications, trackers require manual tuning of covariances, often causing mismatch to the true underlying distribution. Depending on the robustness of the tracker, this can lead to severe reduction of estimation performance. 
To evaluate this, we compared the selected \ac{eot} algorithms on the moderate measurement settings with the trajectory used in Section~\ref{sec:simulation:general}. We introduce a ``mismatch factor'', artificially reducing the kinematic process noise used by the tracker by a certain degree. 

\begin{figure}
    \centering
    \includegraphics[width=0.9\columnwidth]{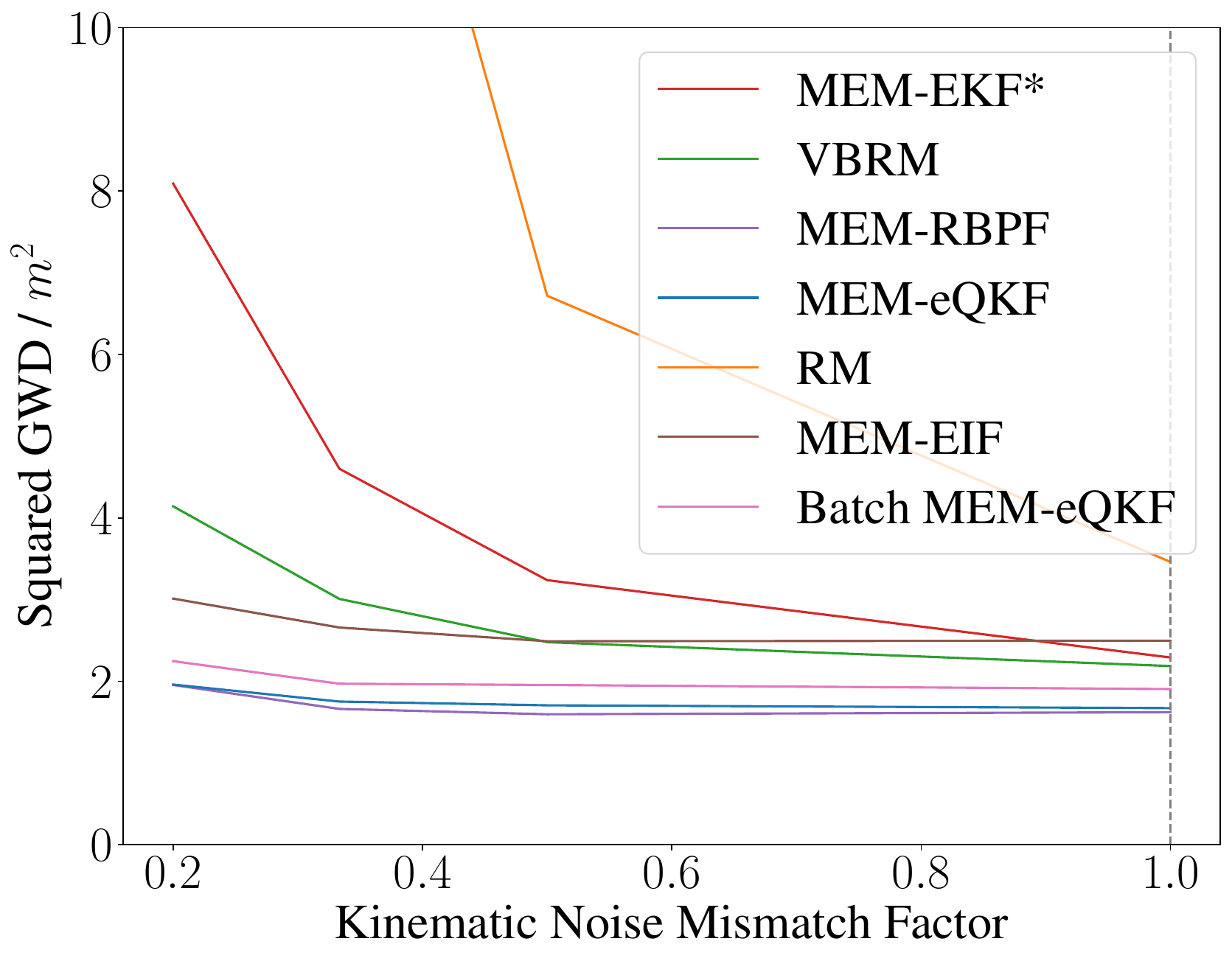}
    \caption{
    Tracking performance for artificially reduced kinematic process noise covariance of the algorithms. For a value of $1.0$ (dashed vertical line), the original covariance is used.
    }
    \label{fig:simulation:mismatch}
\end{figure}

Results are shown in Fig.~\ref{fig:simulation:mismatch}. Four methods are significantly less affected by the factor than the others: The \ac{memeif}, the \ac{memrbpf} and the two variants of the proposed \ac{memeqkf}. 
A key challenge faced by the other reference algorithms is that mismatched motion models cause shape estimation errors, causing a feedback loop of increasing error.

\section{REAL-WORLD EVALUATION}~\label{sec:nuscenes}

\begin{figure}
    \centering
    \includegraphics[width=\columnwidth]{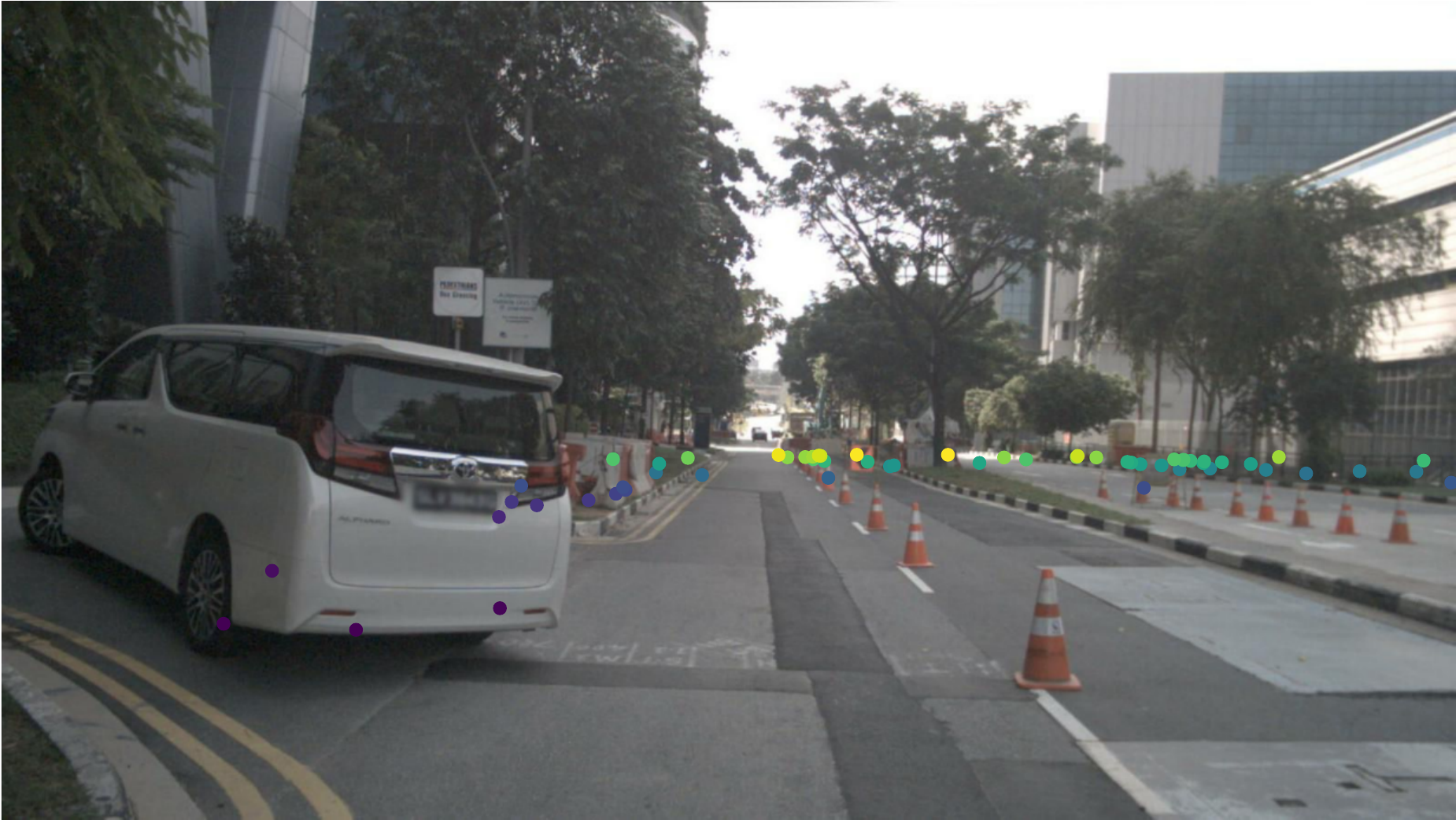}
    \caption{
    Example image from the front camera in the nuScenes~\cite{caesar2020nuscenes} data set, with radar data added as a scatter plot on top. Brighter colors indicate radar points in greater distance. Note the distributed radar detections across the vehicle turning left in front of the recording car.
    }
    \label{fig:nuscenes:data_example}
\end{figure}

In order to assess the applicability of the discussed tracking algorithms to real-world data, we made use of radar data from the nuScenes~\cite{caesar2020nuscenes} data set, an illustration of which is given in Fig.~\ref{fig:nuscenes:data_example}. 
nuScenes contains data from the automotive domain. For each object, consistent annotations throughout each scene are provided as oriented bounding boxes. 

\subsection{Data Set}

We extracted a set of 738 suitable trajectories of individual objects from the nuScenes data set, projected into \ac{bev}. For each object, radar measurements in its vicinity, defined by inflating the ground-truth annotation bounding box by $33\%$, were extracted in each time step. Only sequences belonging to one of the \textit{vehicle} subclasses were extracted. 
The sequences were selected so that each consists of at least ten consecutive annotated time steps.  
Further insight into the distribution of the average number of measurements per step can be found below in Fig.~\ref{fig:nuscenes:error_over_meas}.

The methods' parameters were tuned to the data in a unified manner for all trackers where possible. Some methods require unique parameters, e.g., the number of iterations for \ac{vbrm}. These were tuned individually to yield accurate tracking results on the data set.
The initial covariance matrix for the kinematics was set to $\diag(0.75, 0.75, 1, 1)$, and the kinematic process noise covariance to $\diag(0.01, 0.01, 0.15, 0.15)$.
For the orientation, the initial standard deviation was $10^\circ$ and the process noise variance $0.01$ radians. The variances for the major and minor semi-axes were $1.0$ and $0.05$, respectively, and their process noise variance $10^{-4}$.
The measurement noise was tuned to $\mat{I}_{2\times2}\cdot0.01$. This rather small value stems from the fact that most measurements were indeed observed on the target surface.
For \ac{vbrm}, the parameter $\alpha$ was set to $6$. For \ac{rm}, the initial value for the certainty of the shape estimate was $20$, and the decay of this during the predict step set to a factor of $\exp(-\frac{1}{5})$.
The previously discussed threshold parameter $\psi$ for the axis variance was set to $0.4$ for the \ac{memekf}, and $0.15$ for the batch-based \ac{memeif}.

The prior for all trackers was chosen in the same way: For the kinematics, the mean of the measurements observed in the first time step was used. The velocity vector was set to $\begin{bmatrix}0 & 0\end{bmatrix}^\tp$. 
Eigenvalue decomposition of the sample covariance matrix of the first received measurements was used to acquire an initial orientation estimate. As this can be erroneous for sparse measurements, the corresponding uncertainty needs to be selected large enough.
Finally, for the axis, the mean of all objects in the data set was used as a default value, yielding $\begin{bmatrix}3.93 & 1.26\end{bmatrix}^\tp$ for the semi-axis lengths.

A rectangle is a more accurate description of the target shape than an ellipse. 
Following the underlying assumption that measurement sources are uniformly distributed on the entire object surface, the scaling factor is chosen as $\scalingfactor = \frac{1}{3}$~\cite{steuernagel2024random}, which moment-matches a uniform distribution on a rectangle. Experimental results indeed confirmed that the tracking accuracy of all trackers was noticeably improved by adapting the scaling factor in this manner.

\subsection{Evaluation}
\begin{table}[t]
    \centering
    \caption{
        Comparison of all reference methods with the proposed approach on 738 real-world automotive radar sequences. The median and the mean \acrshort{gws} error of each method are shown.
        Methods are sorted by their mean error.
    }
    \label{tab:nuscenes:quanttable}
    \begin{tabular}{l|cc}
        Method & Median \acrshort{gws} & Mean \acrshort{gws} \\ \hline        
        MEM-RBPF~\cite{steuernagel2025extended}  & 0.260 & 0.336 \\
        MEM-eQKF (ours)  & 0.258 & 0.340 \\
        MEM-EKF*~\cite{yang2019tracking}  & 0.271 & 0.341 \\
        Batch MEM-eQKF (ours)  & 0.263 & 0.348 \\
        VBRM~\cite{tuncer2021random}  & 0.289 & 0.359 \\
        RM~\cite{feldmann2010tracking}  & 0.288 & 0.370 \\
        MEM-EIF~\cite{gramsch2024batch}  & 0.351 & 1375.732 \\
    \end{tabular}
\end{table}
In theory, the \ac{gwd}, used as the evaluation metric in Section~\ref{sec:simulation}, could be employed as the evaluation metric for the nuScenes data, as objects are represented by the very same parameters regardless of whether they represent rectangles or ellipses.
However, the \ac{gwd} is based on the object size. Hence, larger objects produce larger errors. Averaging across a data set with heterogeneous object sizes will then implicitly put more weight on correct estimates of larger objects, skewing the results.

In~\cite{steuernagel2023evaluation}, these issues were discussed in greater detail, and it was suggested to normalize the \ac{gwd} based on the ground truth object size, leading to the so-called \ac{gws}. Even though it is not a true metric, it maintains the desirable attributes of the \ac{gwd}, while eliminating the bias towards large objects when evaluating on the data set at hand. 
In the following, the evaluation results for the nuScenes data will be presented in \ac{gws}.

\subsection{Results}

Compared to the stark difference in performance induced by the maneuvers simulated in Section~\ref{sec:simulation}, the performance of all evaluated trackers on the nuScenes data is rather close. This confirms that all trackers are indeed suitable approaches for radar-based target tracking.

\begin{figure}[t]
    \centering
    \includegraphics[width=0.9\columnwidth]{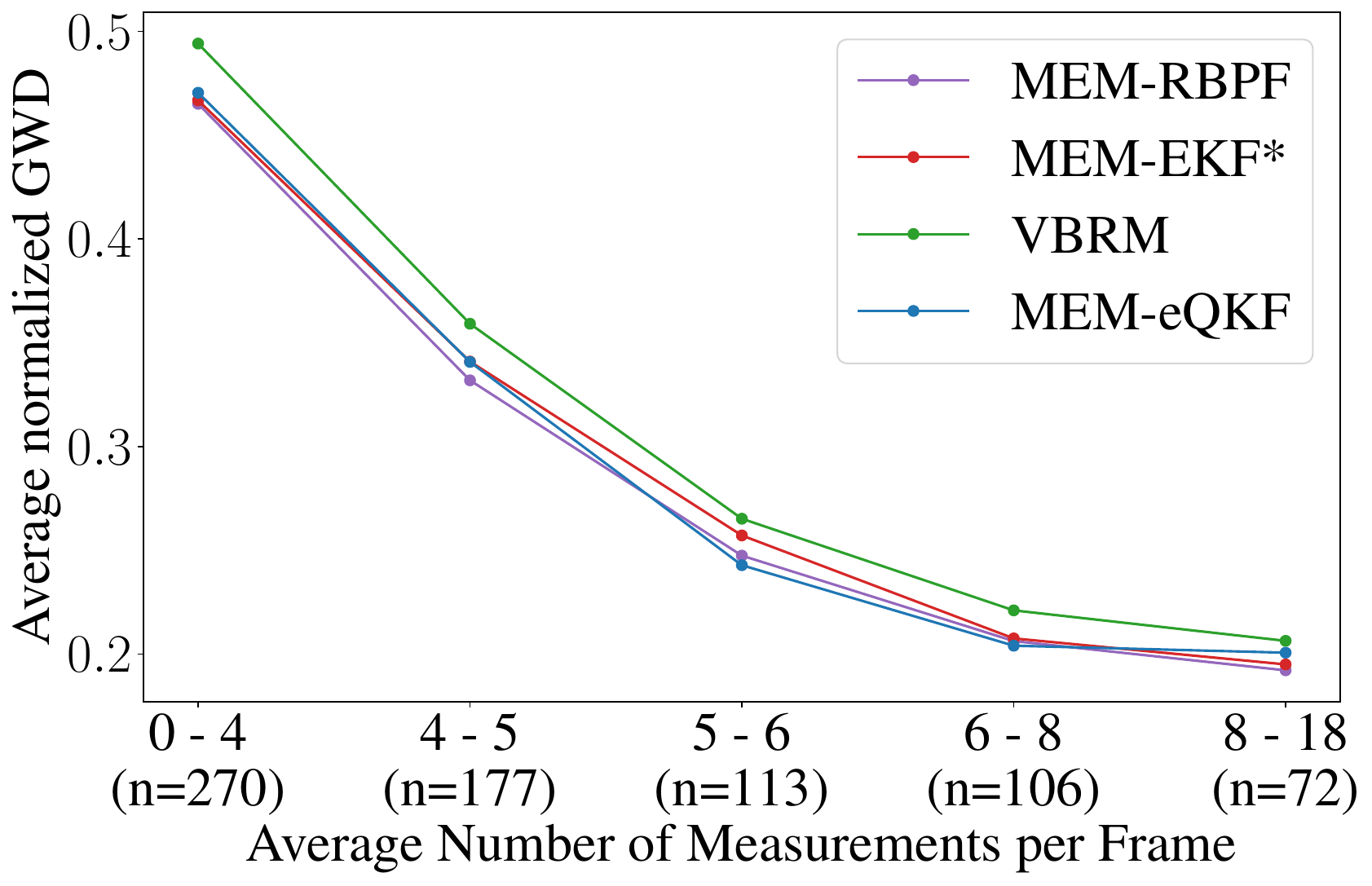}
    \caption{
        Mean error in \acrshort{gws} of each method over the averaged number of measurements in the real-world radar sequences, grouped in bins of relevant measurement numbers.
        Methods that are computed in closed-form are represented with solid lines, others with dashed ones.}
    \label{fig:nuscenes:error_over_meas}
\end{figure}

Nevertheless, when evaluated across hundreds of trajectories, differences in estimation quality are visible.
In Table~\ref{tab:nuscenes:quanttable}, a comparison of each of the reference trackers with the proposed method is presented based on their mean and median \ac{gws} errors.
Due to individual sequences on which no tracker performs well, the mean errors are generally noticeably higher than the median errors. 
The \ac{memeif} yields generally unstable results on this data. It diverges on individual sequences, which causes the mean error to be unreasonably high. 
The proposed tracker yields the lowest error in median, surprisingly even outperforming the sampling-based \ac{memrbpf} on this data.

To evaluate the impact that the number of measurements has on the tracking accuracy, we grouped the extracted sequences based on their average number of measurements. To this end, we created bins of roughly equal number of trajectories for different values of mean number of measurements. 
The results are shown in Fig.~\ref{fig:nuscenes:error_over_meas}.
We focus on the tracking algorithms processing measurements sequentially, and leave out the three batch-based ones. Results for these are still included in Table~\ref{tab:nuscenes:quanttable} for completeness.

In general, the \ac{vbrm} yields slightly worse results than the \ac{mem}-based algorithms across all bins. 
The \ac{memeqkf} and \ac{memrbpf} perform similarly, with the \ac{memekf} producing somewhat worse results than both in most bins.

Fig.~\ref{fig:radar_example} presents a qualitative example of tracking results on the nuScenes data set, comparing results of the \ac{memekf} and \ac{memeqkf}. While similarity in behavior is apparent, it can be seen that the proposed method converges faster, yielding overall better results.

\section{CONCLUSION}~\label{sec:conclusion}

In this article, we proposed an elliptical extended object tracker based on the idea of decoupling the kinematics, orientation, and axes estimation. This results in a deterministic, closed-form algorithm based on Kalman filter equations.
Decoupling these three components has been shown to improve the individual estimation performance.
For example, despite mathematical similarities, our tracker more accurately captures the orientation compared to existing closed-form solutions. 
Importantly, this also allowed for the derivation of a filter that represents the semi-axis lengths in log-normal form, eliminating support for negative semi-axes that is present in all directly related state-of-the-art algorithms such as the \ac{memekf}~\cite{yang2019tracking} or the \ac{memrbpf}~\cite{steuernagel2025extended}.

We furthermore proposed a highly efficient batch-based variant of the filter, requiring significantly less computational load. Nevertheless, this batch-based variant yields competitive results, even outperforming state-of-the-art algorithms that sequentially process all measurements one-by-one.

The results of the method demonstrate a significant advance in estimation accuracy. Our filter achieves comparable results to the best sampling-based reference method, outperforming closed-form and even optimization-based methods.
This was demonstrated on hundreds of sequences of real-world  automotive radar data, and further supported by a comprehensive simulation study where varying measurement conditions were tested.

In theory, the decoupled framework is suboptimal. A joint treatment of the entire state could capture and model correlations between state components explicitly. 
A key example are softly trajectory-aligned objects, where there are strong correlations between shape and velocity orientation, but differences still exist. 
Nevertheless, the results in this paper have shown that the approximations necessary for existing algorithms to treat the problem jointly outweigh the approximations necessary for the factorized approach.

For future work, integration into a multi-object tracker could be of interest.
Furthermore, results on the radar data can be further improved by several means. Improved accounting for specifics of the measurement noise could be achieved by employing, e.g., an \ac{ukf}~\cite{julier1997new}. 
Improved motion models, e.g., via incorporation of an \ac{imm}~\cite{mazor2002interacting}, can also be studied. 
Here, the proposed method has clear advantages over many existing \ac{eot} algorithms due to the modularity arising from the decoupled approach. Replacing the kinematic estimator with existing nonlinear complex motion models is straightforward, and no further derivations are necessary.
In this context, incorporating correlations between shape and velocity orientation by explicitly incorporating the prior estimate of both into both estimation branches is also of interest.

\begin{appendices}
\section{MOMENT DERIVATION}\label{app:moments}
The following section contains the derivation of the closed-form moments $\EX(\mat{a}^i_k), \mat{C}^{\mat{a}, \mat{a}}_{k, i}$ and $\mat{C}^{\mat{a}, \axisstate}_{k, i}$. Due to the similarity of the estimator, we will build closely on the results presented in~\cite[App. A]{steuernagel2025extended}. The results from~\cite{steuernagel2025extended} are not directly applicable, as they are derived for a model where the semi-axes $\semiaxes$ are assumed Gaussian, instead of the log-normal model using $\gamma$ used here.
    
For brevity, in this section, we will omit the time index of variables. We will use $\sigma^2_{\gamma_i} = (\generalcov^{\axisstate})^{i, i}$, i.e., $\sigma^2_{\gamma_i}$ is used as a shorthand for the variance of the $i$-th log-normal axis representation.
We will also make use of $\nu_i$ being the $i$-th dimension of the \textit{aligned} measurement noise, i.e., with the orientation accounted for.

\subsubsection{Expected Pseudo-Measurement}
For the expected pseudo-measurement, 
\begin{subequations}
\begin{align}
    \begin{split}
        \EX(\mat{a})^{(1)} &= \var(h_1\cdot l_1 + \nu_1)
    \end{split}
    \\*
    \begin{split}
        &= \var(h_1\cdot \exp(\gamma_1)) + \var(\nu_1)
    \end{split}
\end{align}
\end{subequations}
For $\var(h_1\cdot \exp(\gamma_1))$ we get
\begin{subequations}
\begin{align}
    \begin{split}
       &\var(h_1\cdot \exp(\gamma_1))  
    \end{split}
    \\*
    \begin{split}
         =& \EX\left[(h_1\cdot \exp(\gamma_1))^2\right] - \left(\EX\left[h_1\cdot \exp(\gamma_1)\right]\right)^2
    \end{split}
    \\*
    \begin{split}
         =& \EX\left[(h_1\cdot \exp(\gamma_1))^2\right] - \left(\EX\left[h_1\right]\EX\left[\exp(\gamma_1)\right]\right)^2
    \end{split}
    \\*
    \begin{split}
        =& \EX\left[h_1^2 \cdot \exp(\gamma_1)^2\right] 
        \label{eq:app:exp-pseudo:drop-e-h}
    \end{split}
    \\*
    \begin{split}
        =& \EX\left[h_1^2\right] \cdot \EX\left[\exp(\gamma_1)^2\right] 
    \end{split}
    \\*
    \begin{split}
        =& \scalingfactor\cdot\EX\left[\exp(\gamma_1)^2\right] 
        = \scalingfactor \cdot \exp\left(2\cdot\hat{\gamma}_1 + 2\cdot\sigma^2_\gamma\right)
        \label{eq:app:exp-pseudo:final}
    \enspace.
    \end{split}
\end{align}
\end{subequations}
We get~\eqref{eq:app:exp-pseudo:drop-e-h} since $\EX[h_1]=0$, canceling out the entire subtracted term.
For~\eqref{eq:app:exp-pseudo:final}, we again make use of the known expectation of the square of a log-normal random variable as already above in~\eqref{eq:memkf:expected-square}.
Finally, $\var(\nu_1) = \mat{W}^{\theta, (1, 1)}_k$, so the overall result is
\begin{equation}
    \renewcommand*{\arraystretch}{2}
        \EX(\mat{a})^{(1)} = 
         \scalingfactor\cdot
         \EX\left[\exp(\axisstate)^2\right]^{(1)} 
         + \mat{W}^{\theta, (1, 1)}
        \enspace.
    \label{eq:app:exp-pseudo:together}
\end{equation}
Identical results are acquired for the second dimension $\EX(\mat{a})^{(2)}$.

\subsubsection{Cross-covariance}
For the cross-covariance $\mat{C}^{\mat{a}, \axisstate}_{k, i}$ it has been shown~\cite{steuernagel2025extended} that the resulting value is a diagonal matrix. For brevity, we will focus on the first entry, corresponding to $l_1$ and $h_1$. We know from~\cite[Eq. (47c)]{steuernagel2025extended} that
\begin{equation}
    (\mat{C}^{\mat{a}, \semiaxes}_{k, i})^{(1, 1)} = \cov(h_1^2 l_1^2, l_1)
    \enspace.
\end{equation}
Adapting to the log-normal form required for the derivation in this article, we get
\begin{subequations}
\begin{align}
    \begin{split}
        & (\mat{C}^{\mat{a}, \axisstate}_{k, i})^{(1, 1)}= \cov(h_1^2 \exp(\gamma_1)^2, \gamma_1)
    \end{split}
    \\*
    \begin{split}
        =&\,\EX[h_1^2\exp(2\gamma_1)\cdot\gamma_1]- \EX[h_1^2\exp(2\gamma_1)]\EX[\gamma_1]
        \label{eq:app:cc:factor-cov}
    \end{split}
    \\*
    \begin{split}
        =&\,\EX[h_1^2]\EX[\exp(2\gamma_1)\cdot\gamma]- \EX[h_1^2]\EX[\exp(2\gamma_1)]\EX[\gamma]
        \label{eq:app:cc:factor-indep}
    \end{split}
    \\*
    \begin{split}
         =&\,\EX[h_1^2]\Bigl(
         \EX[\exp(2\gamma_1)\cdot\gamma]-\EX[\exp(2\gamma_1)]\EX[\gamma]
         \Bigr)
    \end{split}
    \\*
    \begin{split}
        =&\,\scalingfactor\cdot\cov\bigl(\exp(2\gamma_1), g\bigr)
        \enspace,
        \label{eq:app:cc:factor-cov-back}
    \end{split}
\end{align}
\end{subequations}
by using $\cov(X,Y) = \EX[X\cdot Y]-\EX[X]\cdot\EX[Y]$ for~\eqref{eq:app:cc:factor-cov} and~\eqref{eq:app:cc:factor-cov-back} as well as the independence of the random variables for~\eqref{eq:app:cc:factor-indep}.

To compute $\cov\bigl(\exp(2\gamma_1), g\bigr)$, we make use of Stein's Lemma~\cite{stein1981estimation}, letting $f(\gamma_1)=\exp(2\gamma_1)$ and therefore $f'(\gamma_1)=2\cdot\exp(2\gamma_1)$. From Stein's Lemma we then get 
\begin{subequations}
\begin{align}
    \begin{split}
        & \cov\bigl(\exp(2\gamma_1), g\bigr)
    \end{split}        
    \\*
    \begin{split}
        =& \cov(f(\gamma_1), \gamma_1)
    \end{split}
    \\*
    \begin{split}
        =& \cov(\gamma_1,\gamma_1)\EX[f'(\gamma_1)] = \var(\gamma_1)\EX[f'(g)]
        \label{eq:app:cc:stein-applied}
    \end{split}
    \\*
    \begin{split}
        =& \var(\gamma_1)\cdot\EX[2\cdot\exp(2\gamma_1)]
    \end{split}
    \\*
    \begin{split}
        =& \var(\gamma_1)\cdot2\cdot\EX[\exp(\gamma_1)^2]
    \end{split}
    \\*
    \begin{split}
        =& \sigma^2_{\gamma_1}\cdot2\cdot\exp\left(2\cdot\hat{\gamma}_1 + 2\cdot\sigma^2_\gamma\right)
        \enspace,
        \label{eq:app:cc:stein-final}
    \end{split}
\end{align}
\end{subequations}
where~\eqref{eq:app:cc:stein-final} is again making use of the same standard expectation for the square of a log-normal random variable as above. We get~\eqref{eq:app:cc:stein-applied} from Stein's Lemma, which is applicable since $\gamma_1$ is Gaussian and $f(\cdot)$ is differentiable and finite.
Finally, putting~\eqref{eq:app:cc:stein-final} and~\eqref{eq:app:cc:factor-cov-back} together, we get the overall result of
\begin{equation}
    (\mat{C}^{\mat{a}, \axisstate}_{k, i})^{(1, 1)} = 2\cdot\scalingfactor\cdot\sigma^2_{\gamma_1}\cdot\exp\left(2\cdot\hat{\gamma}_1 + 2\cdot\sigma^2_\gamma\right)
\end{equation}
Corresponding results are acquired for the second diagonal entry of the overall diagonal matrix.

\subsubsection{Pseudo-Measurement Covariance}

The marginal measurement covariance $\mat{C}^{\mat{a}, \mat{a}}_{k, i}$ of the pseudo-measurements can be divided into the diagonal and the off-diagonal elements.
Under the assumptions from Section~\ref{sec:methodology}, the off-diagonal elements only depend on $\mat{A}$, exactly as in ~\cite[Eq. (46)]{steuernagel2025extended}, yielding the same result of
\begin{equation}
    \left(\mat{C}^{\mat{a}, \mat{a}}_{k, i}\right)^{(1,2)} = 2\cdot\left(\mat{W}^{\theta, (1, 2)}\right)^2
\end{equation}
and analogously for $\left(\mat{C}^{\mat{a}, \mat{a}}_{k, i}\right)^{(2, 1)}$.

For the diagonal elements, the derivation from~\cite{steuernagel2025extended} can not be used to the log-normal semi-axis distribution used here.
Again, we derive the result for an individual main diagonal element, as the derivation is equivalent for the two. First of all, we get
\begin{subequations}
\begin{align}
    \begin{split}
        \left(\mat{C}^{\mat{a}, \mat{a}}_{k, i}\right)^{(1,1)} = \var\left[(h_1\cdot l_1 + \nu_1)^2\right]
    \end{split}
    \\
    \begin{split}
        &= \EX\left[(h_1\cdot l_1 + \nu_1)^4\right] - \left(\EX\left[(h_1\cdot l_1 + \nu_1)^2\right]\right)^2
    \end{split}
    \\
    \begin{split}
        &= \EX\left[(h_1\cdot l_1 + \nu_1)^4\right] - \left(\EX(\mat{a})^{(1)}\right)^2
    \enspace,
    \end{split}
\end{align}
\end{subequations}
where the second term is already given from~\eqref{eq:app:exp-pseudo:together}. 
For the first term, we use the standard binomial theorem and the independence of the three random variables to acquire
\begin{subequations}
\begin{align}
\begin{split}
      \EX\left[(h_1\cdot l_1 + \nu_1)^4\right]
\end{split} \nonumber
\\
\begin{split}
    &= \EX[(h_1\cdot l_1)^4] + 4\EX[(h_1\cdot l_1)^3]\EX[\nu_1] 
\end{split}
\\
\begin{split}
    &\quad+ 6\EX[(h_1\cdot l_1)^2]\EX[\nu_1^2] + 4\EX[(h_1\cdot l_1)]\EX[\nu_1^3]
\end{split}    \nonumber
\\
\begin{split}
    &\quad+\EX[\nu_1^4]
\end{split}\nonumber
\\
\begin{split}
    &= \EX[(h_1\cdot l_1)^4] + 6\EX[(h_1\cdot l_1)^2]\EX[\nu_1^2] + \EX[\nu_1^4]
    \enspace.
\end{split}
\end{align}
\end{subequations}
Here, we made use of $\EX[\nu_1]  = \EX[\nu_1^3] = 0$ since $\nu$ is a zero-mean Gaussian. 
The terms $\EX[\nu_1^2]$ and $\EX[\nu_1^4]$ are therefore also readily computed making use of the corresponding variance $\mat{W}^{\theta, (1, 1)}$.

\noindent
Additionally, we need
\begin{subequations}
    \begin{align}
        \begin{split}
            \EX[(h_1\cdot l_1)^4] &= \EX[h_1^4]\cdot\EX[l_1^4]
        \end{split}
        \\
        \begin{split}
            &= 3\cdot\scalingfactor^2\cdot\exp\left(4\cdot\hat{\gamma}_1 + 8\cdot\sigma^2_\gamma\right)
        \enspace,
        \end{split}
    \end{align}
\end{subequations}
using the standard moment generating function of the log-normal distribution $l_1$ which gives us ${\EX[\exp(\eta\cdot\gamma)] = \exp(\eta\hat{\gamma}+\eta^2\sigma^2_{\gamma}\frac{1}{2})}$ (with $\eta=4$ here) and the fourth-order central moment for the Gaussian $h_1$. The same but using the second order ($\eta=2$) can be applied to acquire
\begin{equation}
    \EX[(h_1\cdot l_1)^2] = \scalingfactor\cdot\exp\left(2\cdot\hat{\gamma}_1 + 2\cdot\sigma^2_\gamma\right)
    \enspace.
\end{equation}
Finally, putting everything together, we arrive at
\begin{subequations}
\begin{align}
    \begin{split}
        \left(\mat{C}^{\mat{a}, \mat{a}}_{k, i}\right)^{(1,1)}
    \end{split}\nonumber
    \\     
    \begin{split}
        &= \EX\left[(h_1\cdot l_1 + \nu_1)^4\right] - \left(\EX(\mat{a})^{(1)}\right)^2
    \enspace,
    \end{split}
    \\
    \begin{split}
        &= 3\cdot\scalingfactor^2\cdot\exp\left(4\cdot\hat{\gamma}_1 + 8\cdot\sigma^2_\gamma\right)
    \end{split} 
    \\
    \begin{split}
        &\quad+ 6\cdot\scalingfactor\cdot\exp\left(2\cdot\hat{\gamma}_1 + 2\cdot\sigma^2_\gamma\right)\cdot\mat{W}^{\theta, (1, 1)}
    \end{split}\nonumber
        \\
    \begin{split}
        &\quad+ 3\cdot\left(\mat{W}^{\theta, (1, 1)}\right)^2
    \end{split}\nonumber
    \\
    \begin{split}
        &\quad- \left(\EX(\mat{a})^{(1)}\right)^2
        \enspace.
    \end{split}\nonumber
\end{align}
\end{subequations}
The second main diagonal entry $\left(\mat{C}^{\mat{a}, \mat{a}}_{k, i}\right)^{(2,2)}$ can be computed analogously.

\end{appendices}
\bibliographystyle{IEEEtran}
\bibliography{./literature.bib}

@article{stein1981estimation,
  title={Estimation of the mean of a multivariate normal distribution},
  author={Stein, Charles M},
  journal={The annals of Statistics},
  pages={1135--1151},
  year={1981},
  publisher={JSTOR}
}

@article{cao2021automotive,
  title={Automotive radar-based vehicle tracking using data-region association},
  author={Cao, Xiaomeng and Lan, Jian and Li, X Rong and Liu, Yu},
  journal={IEEE Transactions on Intelligent Transportation Systems},
  volume={23},
  number={7},
  pages={8997--9010},
  year={2021},
  publisher={IEEE}
}

@inproceedings{kaulbersch2020assymetric,
  title={Assymetric noise tailoring for vehicle lidar data in extended object tracking},
  author={Kaulbersch, Hauke and Honer, Jens and Baum, Marcus},
  booktitle={2020 IEEE International Conference on Multisensor Fusion and Integration for Intelligent Systems (MFI)},
  pages={191--196},
  year={2020},
  organization={IEEE}
}

@article{xia2021learning,
  title={Learning-based extended object tracking using hierarchical truncation measurement model with automotive radar},
  author={Xia, Yuxuan and Wang, Pu and Berntorp, Karl and Svensson, Lennart and Granstr{\"o}m, Karl and Mansour, Hassan and Boufounos, Petros and Orlik, Philip V},
  journal={IEEE Journal of Selected Topics in Signal Processing},
  volume={15},
  number={4},
  pages={1013--1029},
  year={2021},
  publisher={IEEE}
}

@inproceedings{steuernagel2025decoupled,
  title={Decoupled Quadratic Filter for Extended Object Tracking using Multiplicative Noise},
  author={Steuernagel, Simon and Baum, Marcus},
  booktitle={2025 Sensor Data Fusion: Trends, Solutions, Applications (SDF)},
  pages={1--6},
  year={2025},
  organization={IEEE},
}

@article{mazor2002interacting,
  title={Interacting multiple model methods in target tracking: a survey},
  author={Mazor, Efim and Averbuch, Amir and Bar-Shalom, Yakov and Dayan, Joshua},
  journal={IEEE Transactions on Aerospace and Electronic Systems},
  volume={34},
  number={1},
  pages={103--123},
  year={2002},
  publisher={IEEE}
}

@inproceedings{julier1997new,
  title={New extension of the {Kalman} filter to nonlinear systems},
  author={Julier, Simon J and Uhlmann, Jeffrey K},
  booktitle={Signal Processing, Sensor Fusion, and Target Recognition VI},
  volume={3068},
  pages={182--193},
  year={1997},
  organization={Spie}
}

@inproceedings{steuernagel2024random,
  title={Random matrix-based tracking of rectangular extended objects with contour measurements},
  author={Steuernagel, Simon and Thormann, Kolja and Baum, Marcus},
  booktitle={2024 27th International Conference on Information Fusion (FUSION)},
  pages={1--8},
  year={2024},
  organization={IEEE}
}

@article{li2023tracking,
  title={Tracking of elliptical object with unknown but fixed lengths of axes},
  author={Li, Mingkai and Lan, Jian and Li, X Rong},
  journal={IEEE Transactions on Aerospace and Electronic Systems},
  volume={59},
  number={5},
  pages={6518--6533},
  year={2023},
  publisher={IEEE}
}

@article{lan2023extended,
  title={Extended object tracking using random matrix with extension-dependent measurement numbers},
  author={Lan, Jian},
  journal={IEEE Transactions on Aerospace and Electronic Systems},
  volume={59},
  number={4},
  pages={4464--4477},
  year={2023},
  publisher={IEEE}
}

@article{lan2017tracking,
  title={Tracking of extended object or target group using random matrix: New model and approach},
  author={Lan, Jian and Li, X Rong},
  journal={IEEE Transactions on Aerospace and Electronic Systems},
  volume={52},
  number={6},
  pages={2973--2989},
  year={2017},
  publisher={IEEE}
}

@article{pei2025constrained,
  title={Constrained Extended Target Tracking With Random Matrix and Approximate Projection},
  author={Pei, Shiqi and Wang, Zhen and Li, Ruiyuan and Liu, Jun and Li, Pin and Chen, Chang and Chen, Weidong},
  journal={IEEE Sensors Journal},
  year={2025},
  publisher={IEEE}
}

@article{zheng2025extended,
  title={Extended Object Tracking with Inaccurate Heavy-Tailed Noises},
  author={Zheng, Xiangfei and Zhang, Yujie and Wu, Sunyong and Li, Hongwei},
  journal={IEEE Transactions on Instrumentation and Measurement},
  year={2025},
  publisher={IEEE}
}

@article{zhang2025adaptive,
  title={An Adaptive Multivariate Approach to Dynamic Group Target Tracking Using Variational Inference},
  author={Zhang, Jichuan and Hu, Cheng and Wang, Rui and Jiang, Qi and Shi, Mengxin and Xu, Liang and Tian, Weiming},
  journal={IEEE Transactions on Aerospace and Electronic Systems},
  year={2025},
  publisher={IEEE}
}

@inproceedings{granstrom2016gamma,
  title={Gamma {Gaussian} inverse-{Wishart} {Poisson} multi-{Bernoulli} filter for extended target tracking},
  author={Granstr{\"o}m, Karl and Fatemi, Maryam and Svensson, Lennart},
  booktitle={2016 19th International Conference on Information Fusion (FUSION)},
  pages={893--900},
  year={2016},
  organization={IEEE}
}

@article{wieneke2012pmht,
  title={A {PMHT} approach for extended objects and object groups},
  author={Wieneke, Monika and Koch, Wolfgang},
  journal={IEEE Transactions on Aerospace and Electronic Systems},
  volume={48},
  number={3},
  pages={2349--2370},
  year={2012},
  publisher={IEEE}
}

@inproceedings{granstroem_baum_2022,
    title={A Tutorial on Multiple Extended Object Tracking},
    doi={10.36227/techrxiv.19115858.v1},
    booktitle={TechRxiv},
    author={Granström, Karl and Baum, Marcus},
    year={2022},
    month={Feb}
}

@inproceedings{caesar2020nuscenes,
  title={{nuScenes}: A multimodal dataset for autonomous driving},
  author={Caesar, Holger and Bankiti, Varun and Lang, Alex H and Vora, Sourabh and Liong, Venice Erin and Xu, Qiang and Krishnan, Anush and Pan, Yu and Baldan, Giancarlo and Beijbom, Oscar},
  booktitle={Proceedings of the IEEE/CVF Conference on Computer Vision and Pattern Recognition},
  pages={11621--11631},
  year={2020}
}

@inproceedings{steuernagel2023evaluation,
  title={Evaluation scores for elliptic extended object tracking considering diverse object sizes},
  author={Steuernagel, Simon and Thormann, Kolja and Baum, Marcus},
  booktitle={2023 26th International Conference on Information Fusion (FUSION)},
  pages={1--7},
  year={2023},
  organization={IEEE}
}

@ARTICLE{steuernagel2025extended,
  author={Steuernagel, Simon and Baum, Marcus},
  journal={IEEE Transactions on Signal Processing}, 
  title={Extended Object Tracking by {R}ao-{B}lackwellized Particle Filtering for Orientation Estimation}, 
  year={2025},
  volume={73},
  number={},
  pages={2590-2602},
  doi={10.1109/TSP.2025.3574689}
}

@article{csahin2024random,
  title={Random matrix extended target tracking for trajectory-aligned and drifting targets},
  author={{\c{S}}ahin, Kurtulu{\c{s}} Kerem and Balc{\i}, Ali Emre and {\"O}zkan, Emre},
  journal={IET Radar, Sonar \& Navigation},
  year={2024},
  publisher={Wiley Online Library},
  volume={18},
  number={11}, 
  pages={2247--2263}
}

@article{wen2024velocity,
  title={Velocity-Dependent Orientation Estimation Using Variance Adaptation for Extended Object Tracking},
  author={Wen, Zheng and Lan, Jian and Zheng, Le and Zeng, Tao},
  journal={IEEE Signal Processing Letters},
  year={2024},
  publisher={IEEE}
}

@article{zhang2024extended,
	title = {Extended object tracking using aspect ratio},
	author = {Zhang, Le and Lan, Jian},
	journal = {IEEE Transactions on Signal Processing},
	year = {2024},
  volume={73},
  pages={4193-4207},
	publisher = {IEEE}
}

@article{wang2025extended,
  title={Extended target tracking algorithm based on variational {B}ayes and axis estimation theory},
  author={Wang, Shenghua and Men, Chenkai and Li, Renxian and Yeo, Tat-Soon and He, Pengchao},
  journal={IEEE Transactions on Instrumentation and Measurement},
  year={2025},
  volume={74},
  note = {{A}rt. no. 8507216},
  publisher={IEEE}
}

@article{wen2025extended,
  title={Extended Object Tracking Using an Orientation Vector Based on Constrained Filtering},
  author={Wen, Zheng and Zheng, Le and Zeng, Tao},
  journal={Remote Sensing},
  volume={17},
  number={8},
  pages={1419},
  year={2025},
  publisher={MDPI}
}

@article{zhao2025adaptive,
  title={Adaptive Elliptical Extended Object Tracking Based on Deep Reinforcement Learning},
  author={Zhao, Ziwen and Chen, Hui and Zhang, Wenxu},
  journal={Measurement Science and Technology},
  volume={36},
  number={5},
  year={2025}
}

@article{wahlstrom2015extended,
  title={Extended Target Tracking Using {G}aussian Processes},
  author={Wahlstr{\"o}m, Niklas and {\"O}zkan, Emre},
  journal={IEEE Transactions on Signal Processing},
  volume={63},
  number={16},
  pages={4165--4178},
  year={2015}
}

@article{baum2014extended,
  title={Extended object tracking with random hypersurface models},
  author={Baum, Marcus and Hanebeck, Uwe D},
  journal={IEEE Transactions on Aerospace and Electronic Systems},
  volume={50},
  number={1},
  pages={149--159},
  year={2014},
  publisher={IEEE}
}

@inproceedings{tesori2023lomem,
  title={{L:OMEM}-A fast filter to track maneuvering extended objects},
  author={Tesori, Matteo and Battistelli, Giorgio and Chisci, Luigi and Farina, Alfonso},
  booktitle={2023 26th International Conference on Information Fusion (FUSION)},
  pages={1--8},
  year={2023},
  organization={IEEE}
}

@INPROCEEDINGS{govaers2019independent,
  author={Govaers, Felix},
  booktitle={2019 Sensor Data Fusion: Trends, Solutions, Applications (SDF)}, 
  title={On Independent Axes Estimation for Extended Target Tracking}, 
  year={2019},
  volume={},
  number={},
  pages={1-6},
  doi={10.1109/SDF.2019.8916660}}

@inproceedings{yang2020marginal,
  title={Marginal association probabilities for multiple extended objects without enumeration of measurement partitions},
  author={Yang, Shishan and Wolf, Laura M and Baum, Marcus},
  booktitle={2020 IEEE 23rd International Conference on Information Fusion (FUSION)},
  pages={1--8},
  year={2020},
  organization={IEEE}
}

@inproceedings{kumru2024tracking,
  title={Tracking Arbitrarily Shaped Extended Objects Using {G}aussian Processes},
  author={Kumru, Murat and {\"O}zkan, Emre},
  booktitle={2024 27th International Conference on Information Fusion (FUSION)},
  pages={1--8},
  year={2024},
  organization={IEEE}
}

@article{feldmann2012comments,
  title={Comments on "{B}ayesian approach to extended object and cluster tracking using random matrices"},
  author={Feldmann, Michael and Koch, Wolfgang},
  journal={IEEE Transactions on Aerospace and Electronic Systems},
  volume={48},
  number={2},
  pages={1687--1693},
  year={2012},
  publisher={IEEE}
}

@article{koch2008bayesian,
  title={Bayesian approach to extended object and cluster tracking using random matrices},
  author={Koch, Johann Wolfgang},
  journal={IEEE Transactions on Aerospace and Electronic Systems},
  volume={44},
  number={3},
  pages={1042--1059},
  year={2008},
  publisher={IEEE}
}

@inproceedings{yang2016metrics,
  title={Metrics for performance evaluation of elliptic extended object tracking methods},
  author={Yang, Shishan and Baum, Marcus and Granstr{\"o}m, Karl},
  booktitle={2016 IEEE International Conference on Multisensor Fusion and Integration for Intelligent Systems (MFI)},
  pages={523--528},
  year={2016},
  organization={IEEE}
}

@Article{Givens1984,
  Title                    = {A Class of {Wasserstein} Metrics for Probability Distributions},
  Author                   = {Clark R. Givens and Rae Michael Shortt},
  Journal                  = {The Michigan Mathematical Journal},
  Year                     = {1984},
  Number                   = {2},
  Pages                    = {231-240},
  Volume                   = {31},
  doi                      = {10.1307/mmj/1029003026}
}

@article{tuncer2021random,
  title={Random Matrix Based Extended Target Tracking with Orientation: A New Model and Inference},
  author={Tuncer, Barkin and {\"O}zkan, Emre},
  journal={IEEE Transactions on Signal Processing},
  volume={69},
  pages={1910--1923},
  year={2021},
}

@Article{feldmann2010tracking,
  Title                    = {Tracking of Extended Objects and Group Targets using Random Matrices},
  Author                   = {Feldmann, Michael and Fr{\"{a}}nken, Dietrich and Koch, Wolfgang},
  Journal                  = {IEEE Transactions on Signal Processing},
  Year                     = {2011},
  Number                   = {4},
  Pages                    = {1409--1420},
  Volume                   = {59},
  Doi                      = {10.1109/TSP.2010.2101064},
  ISSN                     = {1053-587X}
}

@inproceedings{gramsch2024batch,
  title={A Batch Update Using Multiplicative Noise Modelling for Extended Object Tracking},
  author={Gramsch, Christian and Yang, Shishan and Alqaderi, Hosam},
  booktitle={2024 27th International Conference on Information Fusion (FUSION)},
  pages={1--8},
  year={2024},
  organization={IEEE}
}

@article{yang2019tracking,
  title={Tracking the orientation and axes lengths of an elliptical extended object},
  author={Yang, Shishan and Baum, Marcus},
  journal={IEEE Transactions on Signal Processing},
  volume={67},
  number={18},
  pages={4720--4729},
  year={2019},
  publisher={IEEE}
}

@inproceedings{baum2012modeling,
  title={Modeling the target extent with multiplicative noise},
  author={Baum, Marcus and Faion, Florian and Hanebeck, Uwe D},
  booktitle={2012 15th International Conference on Information Fusion},
  pages={2406--2412},
  year={2012},
  organization={IEEE}
}

\end{document}